\documentclass[trackchanges,twocolumn]{aastex701}

\expandafter\let\csname ver@longtable.sty\endcsname\relax
\let\longtable\undefined
\let\endlongtable\undefined
\usepackage[T1]{fontenc}
\usepackage[utf8]{inputenc}
\usepackage{subcaption}
\usepackage{amssymb}
\usepackage{amsmath}
\usepackage{empheq}
\usepackage{listings}
\usepackage{graphicx}
\usepackage{cancel}
\usepackage{esint}
\usepackage{xcolor}
\usepackage{hyperref}

\begin{document}

\title{Feeding the Circumplanetary Disk: 3D Simulations of Dust Filtration and Accretion in PDS~70~c}

\author[0009-0003-8984-2094]{Charles H. Gardner}
\affiliation{Department of Physics and Astronomy, Rice University,
Houston, TX 77005, USA}
\affiliation{Rice Space Institute, Rice University, Houston, TX 77005, USA}
\affiliation{Theoretical Division, Los Alamos National Laboratory, Los Alamos, NM 87545, USA}
\email{cg75@rice.edu}

\author[0000-0001-8061-2207]{Andrea Isella}
\affiliation{Department of Physics and Astronomy, Rice University,
Houston, TX 77005, USA}
\affiliation{Rice Space Institute, Rice University, Houston, TX 77005, USA}
\email{isella@rice.edu}

\author[0000-0003-3556-6568]{Hui Li}
\affiliation{Theoretical Division, Los Alamos National Laboratory, Los Alamos, NM 87545, USA}
\email{hli@lanl.gov}

\author[0000-0002-4142-3080]{Shengtai Li}
\affiliation{Theoretical Division, Los Alamos National Laboratory, Los Alamos, NM 87545, USA}
\email{sli@lanl.gov}

\author[0000-0002-2064-0801]{Gennaro D'Angelo}
\affiliation{Theoretical Division, Los Alamos National Laboratory, Los Alamos, NM 87545, USA}
\email{gennaro@lanl.gov}

\author[0000-0001-8291-2625]{Adam M. Dempsey}
\affiliation{X-Computational Physics Divsion, Los Alamos National Laboratory, Los Alamos, NM 87545, USA}
\email{adempsey@lanl.gov}

\begin{abstract}
How circumplanetary disks (CPDs) capture and retain solids is central to constraining the timescale for formation of rocky satellites and interpreting submillimeter continuum observations around giant planets. 
Here, we investigate the transport of gas and dust from the circumstellar disk into the CPD, using PDS~70 as our fiducial example. 
Based on observation-driven parameters, we perform high-resolution 3D adaptive mesh refinement (AMR) hydrodynamic simulations including a multifluid dust component to study dust accretion onto planets with masses of 1 $M_\mathrm{J}$ and 2.5 $M_\mathrm{J}$. 
We find that the pressure maximum at the gap edge imposes strong, size-dependent dust filtration, drastically lowering the solid content of the accreting flow. 
The net dust-to-gas mass ratio of the material accreting onto the planet is reduced by roughly two orders of magnitude relative to the outer disk. 
Only small grains ($\lesssim 61 \mu$m for the 1 $M_\mathrm{J}$ case and $\lesssim 10 \mu$m for the 2.5 $M_\mathrm{J}$ case) are able to accrete efficiently onto the CPD. 
Despite this filtering, we show that a continuous inflow of small grains can still deliver sufficient mass to build the observed PDS~70~c CPD or a Galilean-like satellite system within a few million years. 
Because more massive planets more effectively prevent the accretion of large grains, the dust that reaches the CPD is dominated by small particles with low millimeter-wave opacities. 
Consequently, in the absence of grain growth, interpreting millimeter continuum measurements of CPDs around massive giant planets may require invoking larger total dust masses than typically assumed.
\end{abstract}

\section{Introduction}

How a circumplanetary disk (CPD) acquires and retains solid material is central to resolving two major astrophysical questions: the timescale for rocky satellite formation and the interpretation of observational signatures, such as (sub)millimeter dust continuum emission. 
Previous studies indicate that once a giant planet carves a gap within the circumstellar disk (CSD), a gas pressure maximum forms near the outer gap edge, effectively trapping most solid particles \citep{Whipple+1972, Weidenschilling+1977, Paardekooper+2004, Paardekooper+2006, Rice+2006, Pinilla+2012a, Pinilla+2012b, Zhu2012, Weber2018, Karlin2023}. 
If little to no dust flows across the gap after it opens, the available solid reservoir is strictly limited to the dust mass already present in the CPD at the end of the gap-opening phase. 
In this ``dust isolation'' scenario, both the timescale for forming satellites (or satelliteimals) and the associated dust continuum emission are entirely dictated by the CPD's ability to retain this initial reservoir.

However, the dust isolation framework encounters significant theoretical and observational challenges. 
First, the initial CPD must be massive enough to supply the entire solid inventory required to build satellites. 
For the Jovian system, at least $10^{-4}$ Jupiter masses of solids are necessary to account for the Galilean moons. Assuming a standard primordial gas-to-dust ratio of 100, this requires an initial CPD gas mass exceeding 0.01 $M_\mathrm{J}$. 
This requirement conflicts with the widely accepted ``gas-starved'' model \citep{Canup2002}, which posits that Jupiter's satellites formed within a low-mass CPD ($\sim 10^{-5}$ $M_\mathrm{J}$) continuously replenished with gas and dust over $10^{5}$ to $10^{6}$ years. 
Second, if the CPD lacks internal dust traps, primordial grains would likely undergo rapid radial drift and be accreted by the central planet, forcing an unrealistically short timescale for satellite formation. 
Such rapid solid depletion also contradicts ALMA observations of the dust continuum around PDS~70~c, which reveal that its CPD hosts at least $2\times10^{-5}$ $M_\mathrm{J}$ of small ($a < 1$ mm) dust grains \citep{Isella+2019, Benisty+2021}.
This suggests that CPDs either retain solids for millions of years or are continuously resupplied from the CSD.

An alternative “feeding” scenario proposes that dust slowly escapes from traps located at the outer boundaries of CSD gaps and is then accreted onto the CPD. 
At present, \citet{Karlin2023} and \citet{Szulagyi2022} (hereafter K23 and S22) provide the most extensive studies of dust infall onto CPDs. 
Yet, their findings diverge on the resulting dust content of these disks. 
S22 argues that CPDs can build up substantial reservoirs of millimeter-sized grains, while K23 reports the contrary. 
As K23 emphasizes, this inconsistency arises from fundamentally different methodologies for modeling dust dynamics, particularly in the assumed dust mass distribution and the treatment of diffusion.

Numerical models of dust transport in CSDs rely strongly on the assumed dust Stokes number (St), which sets the degree of hydrodynamic coupling between solid particles and the gas and thus controls the effectiveness of dust trapping. 
Since the Stokes number scales inversely with the gas surface density ($\text{St} \propto 1/\Sigma_\mathrm{g}$), the resulting dust dynamics are extremely sensitive to the adopted gas profile. 
Both S22 and K23 use $\Sigma_\mathrm{g} > 10$ g cm$^{-2}$, corresponding to $\textrm{St}<1$ even for cm-size grains.  
However, earlier modeling of ALMA observations of PDS 70 infers a gas surface density of $\sim$1 g cm$^{-2}$ at the outer edge of the PDS~70~c gap \citep{Keppler2019, Bae2019}, more than an order of magnitude below the values assumed by S22 and K23. 
Such a large discrepancy in gas density can significantly modify the predicted dust distribution and radial drift throughout the disk. 
Consequently, any robust comparison between numerical models and observations must be based on disk parameters that are directly constrained by the data. 
With an abundance of data and a confirmed millimeter wavelength detection of an embedded planet, PDS 70 is a natural choice for constraining disk simulation parameters.

PDS 70 is a K7 T-Tauri star (0.85 $M_\odot$, 0.35 $L_\odot$, 1.26 $R_\odot$) located in the Upper Centaurus-Lupus association at a distance of 113 pc \citep{Pecaut2016, Keppler2018, GAIA2021}. 
The star's CSD features a $\sim$60 au wide cavity, visible in both dust continuum and molecular line imaging \citep{Long2018, Keppler2019, Facchini+2021}. 
The cavity hosts two young, actively accreting protoplanets that have been detected in both infrared and H$\alpha$ emissions \citep{Keppler2018, Haffert2019, Keppler2019, Christiaens+2019, Close+2025}. 
The two planets, PDS~70~b and c, have orbital radii of about 22 au and 34 au, respectively, and masses between 1-5 $M_\mathrm{J}$. 
Crucially, ALMA observations have also detected submillimeter continuum emission co-located with PDS~70~c, confirming the presence of circumplanetary dust, with a total mass between $(2 - 9) \times10^{-5}$ M$_\mathrm{J}$, depending on the assumed grain size \citep{Isella+2019, Benisty+2021}. 
Evidence for circumplanetary dust and planetary accretion strongly supports the theory that PDS~70~c is surrounded by a CPD.  

Measured planetary accretion rates for the PDS 70 system vary significantly depending on the observational tracer and modeling approach used. 
Continual H$\alpha$ monitoring suggests accretion rates on the order of $10^{-9} - 10^{-8}$ $M_\mathrm{J}$/yr, although the H$\alpha$ fluxes themselves exhibit variability \citep{Close+2025, Close+2025b, Zhou+2025}. 
Conversely, \citet{Choksi+2025} employ a shock-heated planetary atmosphere model to match the observed infrared and submillimeter fluxes of PDS~70~c and find a substantially higher accretion rate of $\sim 10^{-7} - 10^{-6}$ $M_\mathrm{J}$/yr. 
If the PDS~70~c disk is in a steady state, the mass accretion rate onto the planet is equal to the mass accretion rate onto the CPD. 
The fraction of this inflow that is in the form of solids depends on how effectively dust grains are trapped at the cavity edge. 
For instance, if mm-sized grains are efficiently retained in the CSD, as proposed by K23, the CPD will be relatively dust-poor compared to the gas and will contain predominantly small grains. 
Consequently, any viable model of the PDS 70 system must also account for and reproduce the observed range of planetary mass accretion rates.
The overarching aim of this work is therefore to compute the dust accretion rate onto PDS~70~c after gap opening and to determine whether the outer CSD can sustain the CPD with a sufficient and continuous influx of solid material.


This paper is organized as follows: In Section 2, we detail the numerical methodology of our high-resolution 3D adaptive mesh refinement (AMR) setup used to model the multifluid dust dynamics. 
In Section 3, we present our results, detailing the 3D gas morphology and the size-dependent filtration of solid material accreting onto the circumplanetary disk. 
We discuss the implications of these accretion rates for satellite formation timescales and the detectability of CPDs in Section 4. 
Finally, we summarize our main conclusions in Section 5.


\section{Numerical Methods and Model Setup}

Our primary objective is to investigate the 3D kinematics and size-dependent filtration of dust accreting onto the CPD of PDS 70 c. 
However, high-resolution 3D adaptive mesh refinement (AMR) simulations featuring multiple dust fluids are computationally prohibitive, making them unsuitable for the broad parameter sweeps required to match observational data. 
To overcome this bottleneck, we first employ a suite of 2D hydrodynamic simulations to efficiently derive the baseline parameters of our PDS 70-analog disk.  

Rather than attempting to reproduce every specific feature of the current observations, which depend heavily on assumed dust properties like grain-size distribution, opacity, and temperature, we use these 2D models to target the broad morphological features of the ALMA Band 7 dust continuum: the ring's location, width, and average peak intensity. 
Because these physical features are highly sensitive to the disk temperature, maximum dust size, dust-to-gas mass ratio ($\Pi$), and adopted viscosity, our 2D parameter sweep allows us to isolate an observationally grounded background state.  

To maintain focus on the 3D accretion dynamics, the full description and results of this 2D calibration are detailed in Appendix \ref{sec:app:2D_meth}. 
The converged parameters from this 2D analysis are then directly utilized to initialize the unperturbed disk in our high-resolution 3D AMR simulations, the setup of which is described below. 


\begin{figure*}[!t]
    \centering
    \includegraphics[width=\textwidth]{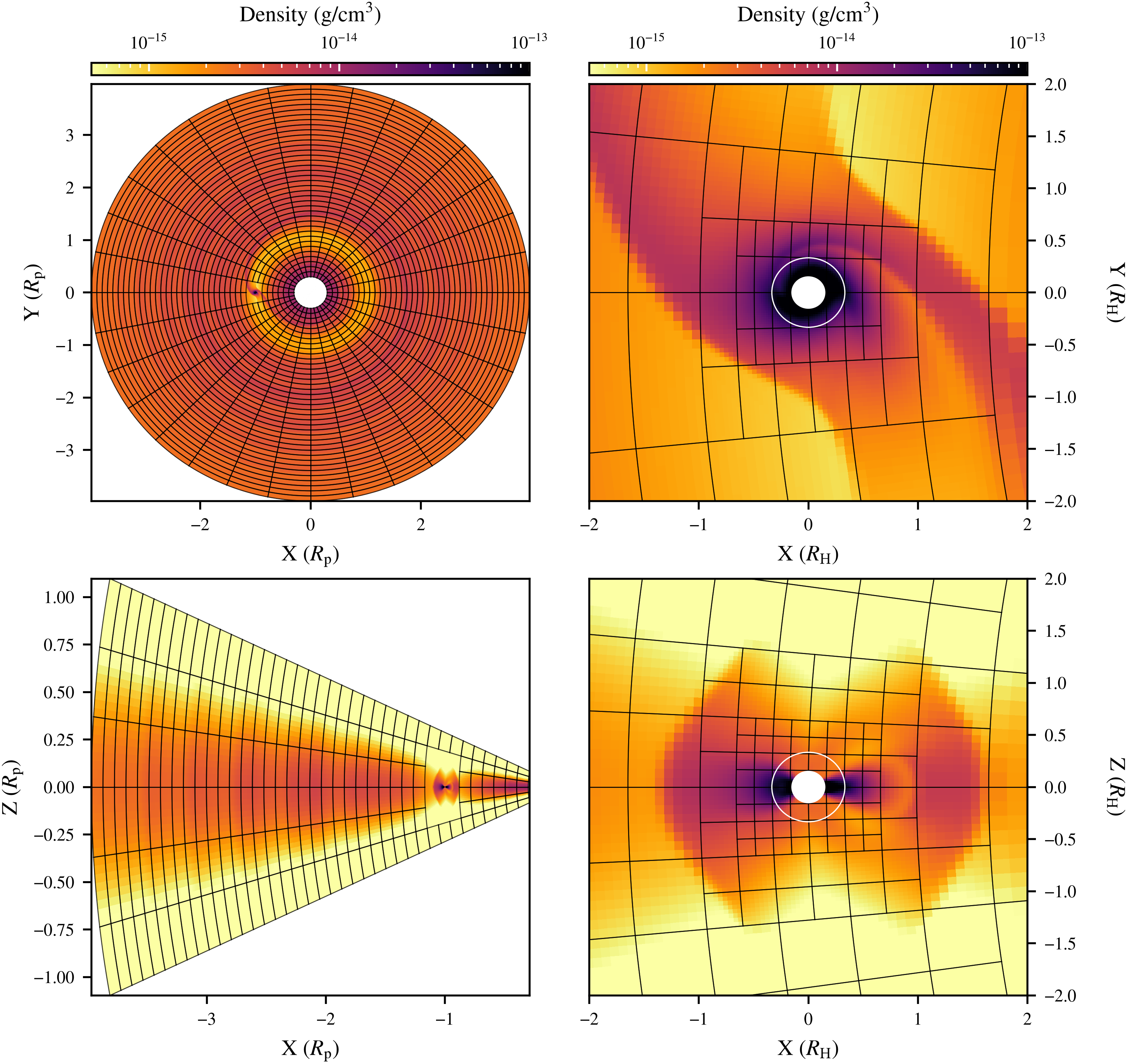}
    \caption{{\it Top:} Slice of gas density along the disk midplane in the 1 $M_\mathrm{J}$ 3D simulations after 1000 orbits at 34 au. The left panel shows the entire disk, while the right shows a zoom-in on $\pm2 R_\mathrm{H}$. The white ring in the zoom-in denotes $1/3 R_\mathrm{H}$, while the white-filled circle denotes the sink radius. {\it Bottom:} The same as the top, but for a vertical slice along $\phi=\pi$ where the planet is located. All panels show the meshblock grid used in the simulations. The right panels show the refined regions. See Section~\ref{sec:3D_Disk} for details about the mesh refinement. }
    \label{fig:grid}
\end{figure*}

\subsection{3D Hydrodynamic AMR Setup}
\label{sec:3D_Disk}

Building on the parameters from our 2D model of PDS 70, we perform global 3D isothermal simulations in spherical polar coordinates.
For this purpose, we employ the Athena++ code with its multifluid dust module, leveraging its adaptive mesh refinement (AMR) capability \citep{Stone2020, HuangBai2022}. 
AMR enables us to locally increase the grid resolution near the planet, ensuring numerical convergence in key regions such as dust traps and circumplanetary disks while keeping the overall computational cost manageable.
Athena++ solves the following continuity and momentum equations for the gas:
\begin{equation}
    \frac{\partial \rho_\mathrm{g}}{\partial t} + \nabla\cdot(\rho_\mathrm{g}\mathbf{v_\mathrm{g}}) = 0,
\end{equation}
\begin{multline}
    \frac{\partial(\rho_\mathrm{g}\mathbf{v}_\mathrm{g})}{\partial t} + \nabla\cdot(\rho_\mathrm{g}\bf{v}_\mathrm{g}\otimes v_\mathrm{g}) = \\
    -\nabla P - \rho_\mathrm{g}\nabla\Phi_\mathrm{G} + \sum_{n=1}^{N_\mathrm{d}}\rho_{\mathrm{d,n}}\frac{\Omega_\mathrm{k}(\mathbf{v_{\mathrm{d,n}}}-\mathbf{v_\mathrm{g}})}{T_{\mathrm{s,n}}} + \nabla\cdot\mathbf{\tau}, 
\end{multline}
where $\rho_\mathrm{g}$, $\mathbf{v_\mathrm{g}}$, $P$, and $\Omega_\mathrm{k}$ denote the gas density, gas velocity, pressure, and Keplerian angular frequency, respectively. 
Since the simulations are isothermal, we omit the energy equation and take the pressure to be $P=\rho_\mathrm{g}c_\mathrm{s}^2$, where $c_\mathrm{s}$ is the isothermal sound speed. 
The term $\Phi_\mathrm{G}$ refers to the gravitational potential due to the star and the planets.
The subscript $n$ on dust quantities in the momentum equation indicates the $n$-th dust species, so the summation term represents the combined back reaction of all dust species on the gas.
$T_{\mathrm{s,n}}$ is a dimensionless stopping time which is defined by \citet{Takeuchi+2002} as:
\begin{equation}
    T_{\mathrm{s,n}} = \sqrt{\frac{\pi}{8}}\frac{\rho_\mathrm{a} a_\mathrm{n}}{\rho_\mathrm{g} H_\mathrm{g}},
\end{equation}
where $a_\mathrm{n}$ is the size of the $n$-th dust grain and $\rho_\mathrm{a}$ is the intrinsic density of a dust grain.
Following the DSHARP mixture, we use a density of 1.675 g/cm$^3$ \citep{Birnstiel+2018}.
$H_\mathrm{g}$ is the gas pressure scale height, defined as:
\begin{equation}
    H_\mathrm{g} = \frac{c_\mathrm{s}}{\Omega_\mathrm{k}},
\end{equation}
$T_{s,n}$ reduces to the familiar Stokes number, St, at the midplane:

\begin{equation}
    \mathrm{St} = \frac{\pi}{2}\frac{\rho_\mathrm{a}a}{\Sigma_\mathrm{g}}.
\end{equation}
$\tau$ represents the viscous stress tensor, which is defined as:

\begin{equation}
    \tau = \rho_\mathrm{g}\nu\left[\nabla\otimes\mathbf{v_\mathrm{g}}+(\nabla\otimes \mathbf{v_\mathrm{g}})^T - \frac{2}{3}(\nabla\cdot\mathbf{v_\mathrm{g}})I\right],
\end{equation}
with $I$ being the identity matrix and $\nu$ being the kinematic viscosity.

The dust continuity and momentum equations of an individual dust species are:

\begin{equation}
       \frac{\partial \rho_\mathrm{d}}{\partial t} + \nabla \cdot(\rho_\mathrm{d}\mathbf{v_\mathrm{d}}) = \nabla\cdot\left(\rho_\mathrm{d}D_\mathrm{d}\nabla\left(\frac{\rho_\mathrm{d}}{\rho_\mathrm{g}}\right)\right),
    \label{eq:dust_cont}
\end{equation}

\begin{multline}
    \frac{\partial (\rho_\mathrm{d}\mathbf{v_\mathrm{d}})}{\partial t} + \nabla\cdot(\rho_\mathrm{d}\mathbf{v_\mathrm{d}}\otimes \mathbf{v_\mathrm{d}}) = \\
    -\rho_\mathrm{d}\nabla\Phi_\mathrm{G}+\rho_\mathrm{d}\frac{\Omega_\mathrm{k}(\mathbf{v_\mathrm{g}-\mathbf{v_\mathrm{d}})}}{T_{s}}+ \nabla\cdot\left(\rho_\mathrm{g} D_\mathrm{d}\nabla\left(\frac{\rho_\mathrm{d}}{\rho_\mathrm{g}}\right)\right)\mathbf{v_\mathrm{d}}.
    \label{eq:dust_mom}
\end{multline}
All dust is modeled as a pressureless fluid, and we again include terms for the gravitational potential and drag.
We also include a diffusion term in the continuity equation and a subsequent correction term to the momentum equation as outlined in \citet{Li+2014}.
The dust diffusion coefficient is set following \citet{Youdin2007} where:
\begin{equation}
    D_\mathrm{d} = \frac{\nu}{1+\text{St}^2}.
\end{equation}
This assumes a gas Schmidt number (Sc) of 1.

The 3D simulations include only a single planet, PDS~70~c. 
This choice is motivated by the stringent timestep constraints imposed by vertical gas and dust motions when simulating massive planets with multiple refinement levels.
Because this computational bottleneck grows with planetary mass, performing long-term integrations for a 5 $M_\mathrm{J}$ planet such as PDS 70 b becomes prohibitively expensive. 
Given that our main goal is to study the transport of dust and gas from the outer disk beyond PDS~70~c onto the circumplanetary disk of PDS 70 c, we conclude that excluding PDS 70 b does not undermine our core scientific aims. 
We run models of PDS 70 c with planet masses of 1 $M_\mathrm{J}$ and 2.5 $M_\mathrm{J}$, mirroring those chosen by \citet{Bae2019}.

Focusing solely on PDS 70 c, we normalize the grid to $r_\mathrm{p}=34$ au and fix the planet's orbit, running the simulations in the corotating frame.
We believe fixing the planet is a sensible choice given that studies in the past have shown the observed PDS 70 configuration to be dynamically stable \citep{Bae2019}.
We also speed up our simulations by running with the FARGO orbital advection algorithm \citep{Masset+2000}.
The radial domain extends from 0.3 to 3.969 $r_\mathrm{p}$, and the azimuthal domain ($\phi$) covers the full $2\pi$. 
The polar extent ($\theta$) is chosen to encompass $\pm4H_\mathrm{g}$ at the normalization radius. 
Our base grid resolution is $304\times48\times512$ ($r\times\theta\times\phi$), giving cells that are approximately cubic, with $\Delta r\simeq r\Delta\theta\simeq r\Delta\phi\simeq0.012 r_\mathrm{p}$.
Throughout the paper, we utilize $r$ to refer to the spherical radius, and $R$ to refer to the cylindrical radius.

The base grid is decomposed into meshblocks of $8\times8\times16$ cells. 
Meshblocks located within half of the planetary Hill radius, $R_\mathrm{H} = r_\mathrm{p} (M_\mathrm{p}/3M_\star)^{1/3}$, are refined by three additional levels, yielding a finest resolution of $\sim1.5\times10^{-3}r_\mathrm{p}$ in all directions. 
For the 1 $M_\mathrm{J}$ and 2.5 $M_\mathrm{J}$ simulations, this corresponds to spatial resolutions of $\sim0.02 R_\mathrm{H}$ and $\sim0.015 R_\mathrm{H}$, respectively.
The resulting grid architecture, including the nested levels of mesh refinement, is visualized in Figure~\ref{fig:grid}.

Our simulations adopt an initial gas density calculated using:
\begin{equation}
\label{eq:rho}
\rho_{\mathrm{g,0}}(R,z) = \frac{\Sigma_{\mathrm{g,0}}(R)}{\sqrt{2\pi}H_\mathrm{g}(R)} e^{-z^2/2H_\mathrm{g}(R)^2},
\end{equation}
where $z=r\cos\theta$ denotes the height above the midplane and the initial surface density profile, $\Sigma_{g,0}$ was chosen to establish a viscous, quasi-steady-state accretion flow through the initial unperturbed disk.
According to the similarity solution for a viscously evolving disk with constant $\nu$, the steady-state surface density corresponds to a Gaussian function with width equal to $\sqrt{2}R_t$.
$R_t$ is the critical radius, i.e., the radius at which the radial velocity of the gas flips from negative (inward motion) to positive \citep[outward motion; see, e.g.,][]{Isella+2009}.
To reach a steady-state accretion profile in the CSD, we assume a disk outer radius $R_{out}\ll R_t$ and, within $R_{out}$, a constant initial gas surface density, $\Sigma_{\mathrm{g,0}}$.
We set $\Sigma_{\mathrm{g,0}}$ to be 1.2 g/cm$^2$ at $R_\mathrm{p}$ following the results of our 2D analysis (see Section \ref{sec:app:2D_res}).

The chosen pressure scale height and temperature profile of the disk also come from our 2D analysis and result in an aspect ratio of:
\begin{equation}
\label{eq:aspect}
    \frac{H_\mathrm{g}}{R}(R) = 0.07\left(\frac{R}{34\text{ au}}\right)^{0.2}.
\end{equation}
However, once the planet is included in the simulations, we modify the temperature of the disk near the planet as \citep{YapingLi2022}
\begin{equation}
\label{eq:planet_temp}
    \tilde T(R,t) =T(R)\times\begin{cases}
        \left( 1+s(t)\times\frac{\epsilon(0.1R_\mathrm{H})^2}{d^2 +\left(0.1R_\mathrm{H}\right)^2} \right)  &\text{ if d $\leq$ $R_\mathrm{H}$}\\
        1 & \text{ if d $>$ $R_\mathrm{H}$},
    \end{cases} 
\end{equation}
where $d$ is the spherical distance from the planet and $\epsilon$ is a scaling factor which we set to 10.
This raises the temperature of the gas surrounding the planet to account for the extra heating produced by the planet’s own emission and accretion. 
At the planet itself, the temperature is increased by a factor of 11 to $\sim330K$.
Although approximate, it reproduces the expected temperature profile of the circumplanetary disk environment around a planet with the inferred properties of PDS 70 c \citep{Wang2021}.
To avoid any thermal instabilities, we smoothly ramp the sound speed near the planet up to this final value with a sine function:
\begin{equation}
s(t) = 
    \begin{cases}
        0 & \text{ if t $<$ $t_0$}\\
        \sin\left(\frac{\pi}{2}\frac{t-t_0}{t_1-t_0}\right) & \text{ if $t_0\leq t<t_1$} \\
        1 & \text{ if $t \geq t_1$},
    \end{cases}
\end{equation}
Where $t_0$ here is the time the temperature increase begins and $t_1$ the time that it ends.
We set $t_0$ to 5 orbits and $t_1$ to 12 orbits.

The initial radial gas velocity is azimuthally symmetric and is obtained from Eq. 11 of \citet{Takeuchi+2002}, with a modification to incorporate constant viscosity:
\begin{equation}
    v_{r, g}(R, \phi, z) = -\frac{\nu}{2R}\left(6p - 2q + 3 + (5q + 9)\left(\frac{z^2}{H(R)^2}\right)\right),
\end{equation}
where $p=-1.2$ is the power-law index of the midplane density profile, and $q=-0.6$ is the power-law index of the temperature profile.
These indices result in a constant $\Sigma_{\mathrm{g,0}}$ disk for the temperature profile introduced in Equation \ref{eq:aspect}.
At the midplane, the radial velocity is directed away from the star while at higher altitudes it points towards the star.
We adopt a kinematic viscosity $\nu=1.2\times10^{16}$ cm$^2$ s$^{-1}$ which corresponds to a \citet{SS+1973} parameter of $\alpha=10^{-2}$ at the location of PDS 70 c and $\alpha=5\times10^{-3}$ at the radius of the dust ring seen in observations. 
The initial azimuthal gas velocity is prescribed as:
\begin{equation}
    v_{\phi,g}^2(R,\phi,z) = \frac{GM_\star R^2}{(R^2+z^2)^{3/2}} + \frac{R}{\rho}\frac{\partial P}{\partial R}.
\end{equation}
The initial polar velocity is chosen such that $v_{\theta,g} = v_{r,g}\cos\theta$, 
ensuring that it remains a small fraction of the radial velocity across most of the simulation domain.

We employ radial viscous boundary conditions so that the disk relaxes toward a viscous steady-state solution. 
Furthermore, we apply wave-damping zones \citep{ValBorro+2006} to prevent the reflection of spiral density waves, launched by the planet, at the inner and outer radial boundaries of the computational domain.
The details of the boundary condition implementation are provided in Appendix~\ref{app:bound_cond}. 

To obtain a realistic 3D simulation of gas and dust dynamics in CPD regions, we must carefully account for the impact of finite grid resolution near the planet. 
In an ideal scenario, if the grid could be refined indefinitely (and all relevant physical processes were included), we could model the gas and dust dynamics down to the planetary surface. 
In practice, however, our computational resources restrict us to resolving the flow only down to $0.15 R_{\mathrm{H}}$. 
Within $\sim10$ grid cells from the planet, the physical reliability of our simulation deteriorates. 
A standard approach to alleviating this issue is to introduce a spherical sink region where material is artificially removed on a time scale $\tau_{\mathrm{s}}$.
Following \cite{Choksi+2023}, we prescribe the removal timescale as $\tau_{\mathrm{s}} = r_{\mathrm{s}}/c_{\mathrm{s}}$, where $r_\mathrm{s} = 0.15 R_\mathrm{H}$ is the sink-sphere radius and $c_{\mathrm{s}}$ is the sound speed at $r_{\mathrm{p}}$.
The gravitational potential of the planet is also softened over $0.15R_\mathrm{H}$ using a cubic spline function.
The removal timescale corresponds to removing gas and dust at a rate:
\begin{equation}
    \frac{\partial\rho}{\partial t}  = -\frac{\rho}{\tau_{\mathrm{s}}} \, .
\end{equation}
We further impose a quadratic radial dependence of the sink timescale within the sink sphere:
\begin{equation}
    \tau_{\mathrm{s}}(d) = \tau_{\mathrm{s}}\left(1 - \left(\frac{d}{r_{\mathrm{s}}}\right)^2\right) \, ,
\end{equation}
such that the maximum removal rate is attained only at the center.

With the adopted AMR setup, the sink sphere is resolved with 10 cells, which is sufficient to apply the torque-free accretion method described in \citet{Dempsey+2020b}.
This sink radius corresponds to roughly half the typical CPD size, $\sim1/3 R_\mathrm{H}$, indicating that our 3D models do not fully resolve the CPD. 
This limitation is acceptable as our primary objective is to investigate how gas and dust are transported from the CSD into the CPD region at a distance of about $1/3$ $R_\mathrm{H}$ from the planet, rather than to assess the ultimate fate of this material (i.e. whether it is retained in the CPD or accretes immediately to the planet).
Assessing that retention efficiency would require resolving the CPD's internal structure, which is beyond the scope of this study and left for future work.
As a summary, we provide a table of all of our normalization parameters in Table \ref{tab:param}.

\begin{deluxetable}{cc}[!t]
\tablehead{
\colhead{Quantity} &
\colhead{Value} 
}
\startdata
$r_\mathrm{p}$ & 34 au \\
$\Sigma_0$ & 1.2 g/cm$^2$ \\
$\rho_0$ &  $1.3\times10^{-13}$ g/cm$^3$\\
$H_\mathrm{g}$/$r_\mathrm{p}$ & 0.07 \\
$T_0$ & 30.2 K\\
$\nu$ & $1.2\times10^{16}$ cm$^2$/s \\
$\alpha_0$ & $10^{-2}$ \\
\enddata
\caption{Notable disk simulation parameters used in our 3D simulations at $r=r_\mathrm{p}$. \label{tab:param}}
\end{deluxetable}

\subsection{Planet Insertion and CPD formation}

\begin{figure}[!t]
    \centering
    \includegraphics[width=1.0\columnwidth]{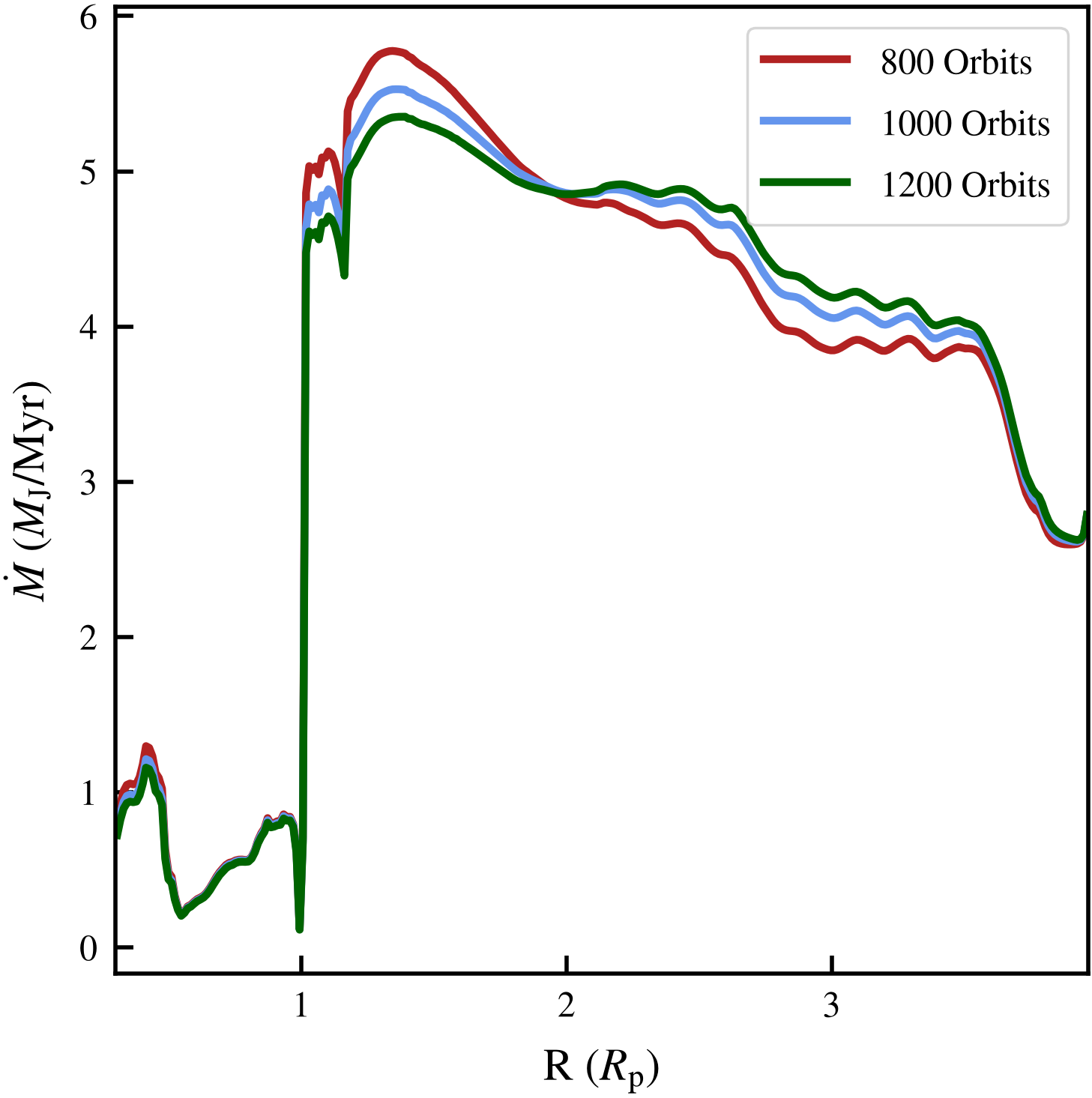}
    \caption{Global disk accretion rate as a function of radial distance from the star. The planet extracts material at 1 $R_\mathrm{p}$, producing a discontinuity between the inner and outer disk. The accretion rate outside the planet’s orbit equals the sum of the material removed by the planet and the accretion rate inside its orbit. The outer disk continues to evolve because it has not yet completed a full viscous timescale. However, as time progresses, the subsequent changes in the $\dot{M}$ profile get smaller and smaller. 
    }
    \label{fig:global_mdot}
\end{figure}

Standard hydrodynamic simulations of planet-disk interactions generally do not self-consistently model the planet's formation. 
Instead, to avoid numerical instabilities, a planetary core (e.g., with an initial mass of 10 Earth masses) is typically inserted into an unperturbed disk, and its mass is increased over several orbits until it reaches the desired final mass. 
Consequently, the initial formation of the CPD is not properly captured, meaning the resulting CPD mass (in both gas and solids) and its early structure may not be physically realistic.

To mitigate these artificial initial conditions, we first run simulations with only gas for 1000 planetary orbits, corresponding to approximately $2\times10^5$ years. 
Over the initial 5 orbits, the planet grows from 0 to its final target mass of either 1 or 2.5 $M_\mathrm{J}$ as follows:
\begin{equation}
    M_\mathrm{p}(t) = M_\mathrm{p}\times\begin{cases}
        0 & \text{if } t < t_0 \\
     \sin^2\left(\frac{\pi}{2}\frac{t - t_0}{t_1 - t_0}\right) & \text{if } t_0 \leq t < t_1 \\
    1 & \text{if } t \geq t_1,
    \end{cases}
    \label{eq:mass_grow}
\end{equation}
In this setup, we choose $t_0$ to correspond to 2 orbits after the start of the simulation and $t_1$ to correspond to 7 orbits, giving a total growth interval of 5 orbits. 
Over the following 993 orbits, the planet opens a gap in the CSD, and a gas-only CPD develops self-consistently.
To determine whether the disk has reached a quasi-steady state, we calculate the mass accretion rate ($\dot{M}$) profile throughout the disk as:
\begin{equation}
    \dot{M}(r)=-\int_{\phi_{min}}^{\phi_{max}}\int_{\theta_{min}}^{\theta_{max}}\rho_\mathrm{g}v_{\mathrm{r,g}}r^2\sin(\theta) d\theta d\phi.
\end{equation}
The negative sign defines $\dot{M}$ so that positive values point towards the star, or planet, and represent material being added to it.
By the conclusion of this gas-only stage, both the CSD and the CPD have relaxed into configurations that exhibit a steady-state $\dot{M}$ profile (see Figure~\ref{fig:global_mdot} for the 1 $M_\mathrm{J}$ model and Figure~\ref{fig:global_mdot_2p5} for the 2.5 $M_\mathrm{J}$ model).

While the viscous timescale at the location of the planet is $\sim3000$ orbits, after 1000 orbits $\dot{M}$ is positive across the entire disk and exhibits minimal temporal variation.
The disk has reached a viscous time interior of $\sim0.6$ $R_\mathrm{p}$ while larger radii have not.
However, Figure \ref{fig:global_mdot} shows that the rate of change of $\dot{M}$ over the whole disk is decreasing, as the changes in the profile from 800 orbits to 1000 orbits are much larger than those from 1000 to 1200.
This motivated our insertion of dust after 1000 orbits.

The discontinuity observed at $R=R_\mathrm{p}$ occurs because material from the outer disk crosses into the CPD and enters the sink sphere surrounding the planet, where it is subsequently removed from the simulation domain. 
Our 2D tests have shown that eventually this $\dot{M}$ profile should flatten on either side of the planet with the difference between the the outer and inner disk being what is removed.
However, this takes on the order of a viscous time, so we instead study the stage where there are some changes in the $\dot{M}$ profile, but the differences between subsequent time steps are getting smaller and smaller.
At this stage, in the 1 $M_\mathrm{J}$ model, the gas mass accretion rate onto the CPD (measured at $1/3$ $R_\mathrm{H}$) is approximately 4 $M_\mathrm{J}$/Myr, and the total CPD gas mass is $1.7\times10^{-4} M_\mathrm{J}$.
Notably, this value falls comfortably within the low-mass range predicted by the gas-starved accretion model of satellite formation \citep{Canup2002}.

Although the total CPD mass decreases slightly over time, the accretion rate at $1/3$ $R_\mathrm{H}$ converges to within a few percent of the sink mass-removal rate, confirming that the system has reached a quasi-steady state. 
By this point, the CPD has had ample time to erase any unphysical artifacts introduced by the planet's rapid, artificial growth. 
Given that the local orbital timescale at $1/3$ $R_\mathrm{H}$ is roughly 22 years, the CPD completes more than $10^4$ orbits during the gas-only initialization phase.
With a viscous timescale around the CPD ($1/3R_\mathrm{H}$) of $\sim 400$ years, this ensures a well-relaxed hydrodynamic CPD structure prior to the introduction of dust.

We also examine the depth of the gap around the planet as a means of assessing whether the simulation has reached a quasi-steady equilibrium.
By 1000 orbits, the gap depth ($\Sigma_{\mathrm{min}}/\Sigma_0$) varies by only a few percent per 100 orbits, having settled to a value of $\sim10\%$.
The same behavior is seen in our 2D simulations: by 1000 orbits of PDS 70 c, the gap depth likewise changes by no more than a few percent per 100 orbits.
In this case, however, the gap settles to an absolute depth of $\sim20\%$.
This difference in absolute depth arises because the 3D simulations include the sink sphere, which continuously removes gas from the gap region, while the 2D simulations do not include planetary accretion.
Both simulations are therefore independently converged, despite reaching different absolute gap depths.

The mass accretion rate onto the CPD and the CPD gas mass for the 2.5 $M_\mathrm{J}$ model are similar (Appendix~\ref{app:2.5mj}).

\subsection{Dust Initialization and Boundary Conditions}
After completing the 1000-orbit gas-only initialization stage, we need to add solid particles to the simulation. 
In a perfect setup with unrestricted computational power, we would continuously feed multiple dust species in at the disk’s outer edge, let them drift radially inward, move through the circumstellar disk, and ultimately be accreted by the CPD. 
In practice, though, the radial drift timescales of small grains are so long that this method becomes computationally impractical.

Instead, we populate the outer CSD with dust, scaling the dust density to the local gas density and assuming a global initial dust-to-gas mass ratio of $\Pi = 0.02$ (see Appendix \ref{sec:app:2D_res}) together with an MRN grain-size distribution \citep{MRN1977}.
Importantly, we confine this initial dust reservoir to radii larger than $1.5~R_\mathrm{p}$ so that the circumplanetary region is completely free of solids at the beginning of the simulation with dust.
This corresponds to a distance of $\sim 7 R_\mathrm{H}$ for our 1 $M_\mathrm{J}$ model and $\sim 5 R_\mathrm{H}$ for our 2.5 $M_\mathrm{J}$ model.
As a result, any dust that ultimately enters the CPD must first cross the gas pressure maximum (i.e., the dust trap), which, as detailed in the following section, is located at 1.5 $R_\mathrm{p}$. 
This spatial configuration separates the different transport processes, allowing us to better understand how dust moves from the outer disk toward the planet.
The dust is initialized in the simulation domain at the velocity of the gas before being allowed to evolve independently.

To maintain a continuous supply of material during the run, the dust density in the outer radial ghost zones is dynamically scaled to the local gas density using our assumed initial dust-to-gas mass ratio. 
Furthermore, we set each component of the dust velocity in these outer ghost zones to equal the local gas velocity. 
This ensures that the dust mass flux ($\dot{M}_\mathrm{d}$) injected into the domain is exactly the gas $\dot{M}$ scaled by the initial dust-to-gas mass ratio, allowing us to strictly govern the amount of solid material supplied by the outer disk. 
All other boundary conditions for the dust species are identical to those implemented for the gas.

The specific dust grain sizes included in the simulations are listed in Table~\ref{tab:mdot} and vary slightly between the 1 $M_\mathrm{J}$ and 2.5 $M_\mathrm{J}$ models. 
This adjustment is made to better sample the relevant range of Stokes (St) parameters as the planetary mass increases, a choice discussed further in the following section. 
The dust mass in each size bin is calculated following an MRN size distribution $n(a)\propto a^{-3.5}$ and constant $\Pi$.
The maximum grain size of the distribution is set to 1 cm following the results of \citet{Sierra+2025} for the PDS 70 ring.
The size of the dust is chosen so that the mass within a bin is split evenly on either side of the dust size that represents the bin (see Appendix \ref{sec:app:dust}).

\section{Results}

\subsection{Global Circumstellar Disk Properties and Dust Trapping in 3D}


\begin{figure*}[!]
    \centering
    \begin{subfigure}[b]{1.0\textwidth}
        \centering
        \includegraphics[width=0.9\textwidth]{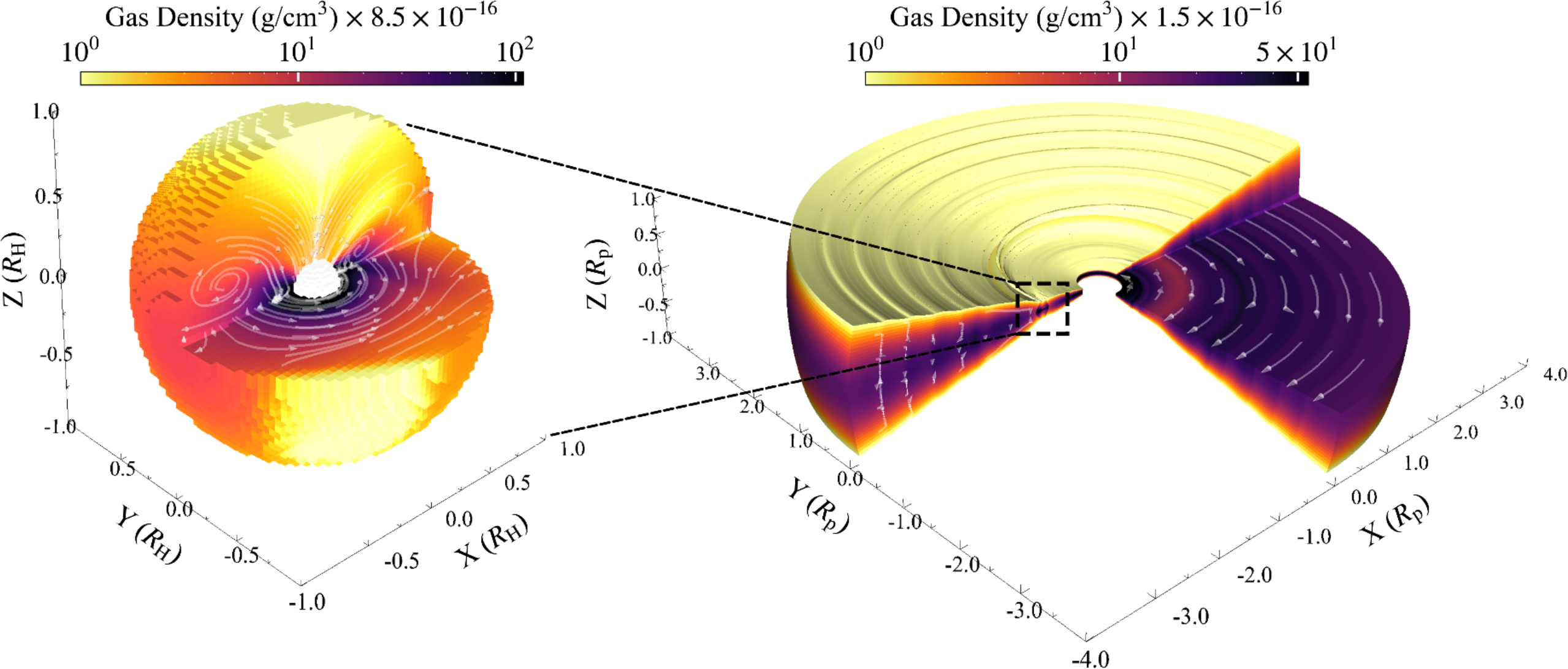}
        \label{fig:sub1}
    \end{subfigure}
    \par\vfill
    \begin{subfigure}[b]{1.0\textwidth}
        \centering
        \includegraphics[width=0.9\textwidth]{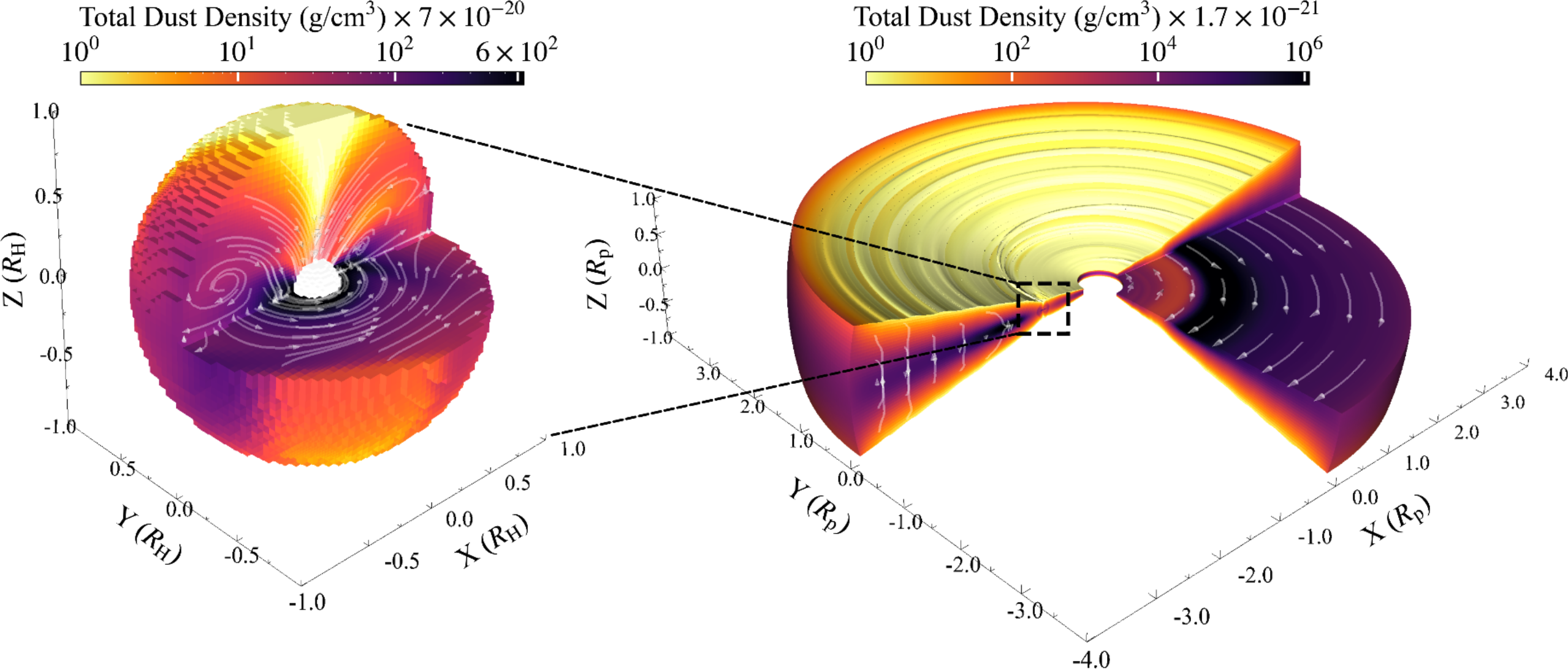}
        \label{fig:sub2}
    \end{subfigure}
    \par\vfill
    \begin{subfigure}[b]{1.0\textwidth}
        \centering
        \includegraphics[width=0.9\textwidth]{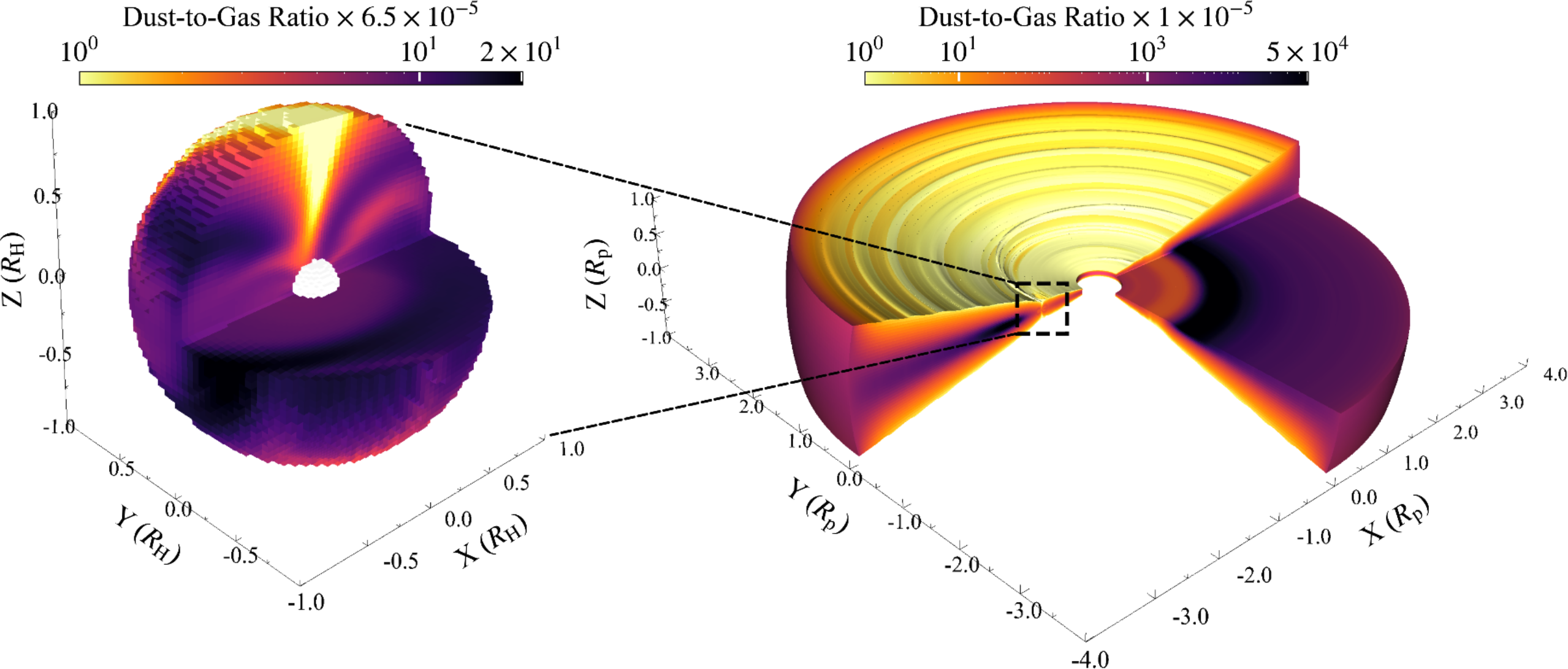}
        \label{fig:sub3}
    \end{subfigure}
    \caption{3D rendering of the 1 $M_\mathrm{J}$ simulation taken 1000 orbits after dust is introduced. The left column has coordinates centered on the planet and in terms of the Hill radius while the right column is centered on the star and is in terms of $R_\mathrm{p}$. In each row, the white sphere within the zoom-in of the inner Hill radius indicates the sink sphere. Each colorbar saturates at the 1st percentile for the minimum and the 99.5 percentile for the maximum, except for the Hill sphere picture of the dust-to-gas mass ratio which saturates at the 99.99 percentile. The minimum of the colorscale is also factorerd out to show the relative range. {\it Top:} Gas density across the disk and inside the Hill sphere. {\it Middle:} As in the top row, but showing the total dust density. In both the gas and total dust density panels, streamlines illustrate the midplane and XZ-plane projections of the total velocity vector. For the total dust velocity, we apply a density-weighted combination of the velocities of all dust species. {\it Bottom:} As above, but displaying the dust-to-gas mass ratio.} 
    \label{fig:combined}
\end{figure*}

Figure~\ref{fig:combined} presents 3D visualizations of the gas density, total dust density, and dust-to-gas mass ratio at the end of our simulation for the 1 $M_\mathrm{J}$ planet case. 
The figure also displays gas and dust streamlines, where the dust streamlines are based on the density-weighted velocity computed across all dust species.  
An analogous figure for the 2.5 $M_\mathrm{J}$ planet model is provided in the Appendix~\ref{app:2.5mj}. 

The planet carves out an almost circular gap in both the gas and dust, extending from 0.9 to 1.1 $R_\mathrm{p}$. 
Outside this gap, the motion of gas and dust is close to Keplerian, with the azimuthal velocity $v_\phi$ dominating the kinematics. 
There are also finite radial and vertical velocity components, which become more pronounced for larger dust grains (with higher Stokes numbers) that are less strongly coupled to the gas, leading to more efficient inward drift and vertical settling.
Consequently, most of the dust has settled toward the midplane and accumulated in a dust pressure trap within the gas ring at the outer boundary of the planet-created gap.

\begin{figure*}[t]
    \centering
    \includegraphics[width=0.8\textwidth]{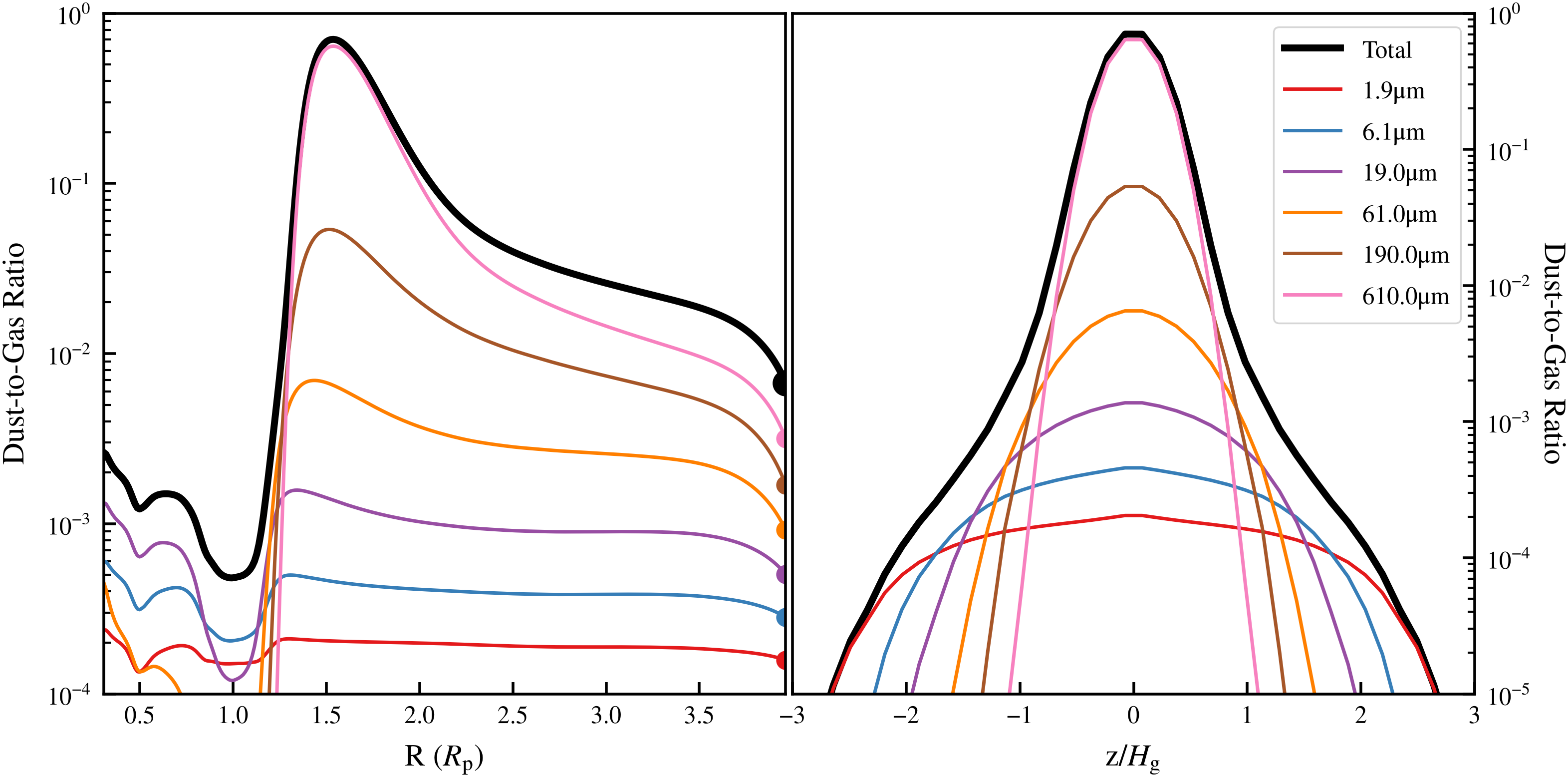}
    \caption{{\it Left:} Dust-to-gas mass ratio of each individual dust species and of the total dust population in the 1 $M_\mathrm{J}$ model, shown as a function of radius in the global disk. Most of the dust mass in the dust trap resides in the largest grains, while only the smallest grains are able to pass through the gap. {\it Right:} Same as in the left panel, but now plotted as a function of height above the midplane in terms of gas scale heights, $H_\mathrm{g}$. The radial location used for this plot and for calculation of $H_\mathrm{g}$ is where the dust-to-gas mass ratio reaches it maximum, R = 1.5 $R_{\mathrm{p}}$. Large grains remain confined to the midplane, whereas small grains extend to higher altitudes in the disk.
    }
    \label{fig:radial_D2G}
\end{figure*}

The left panel of Figure \ref{fig:radial_D2G} shows the radial profile of the midplane dust-to-gas mass ratio $\Pi$ over the full disk, plotted separately for each dust species and for their sum. 
The total dust-to-gas mass ratio exhibits a pronounced dust ring centered at $\sim1.5\,R_\mathrm{p}$ (51 au), coincident with the gas pressure maximum. 
We define the dust trap as the radial region where $\Pi > 0.2$ (i.e., ten times the initial value of $\Pi$). Under this definition, the dust pressure trap extends from roughly 1.4 $R_\mathrm{p}$ to 1.9 $R_\mathrm{p}$.
At these radial locations, the vertically integrated, density weighted pressure is within 10\% of the pressure maximum.
At the center of the trap, $\Pi$ reaches a peak value of $\sim0.7$, corresponding to an enhancement of about 35 times relative to the initial dust-to-gas mass ratio. 
Within the trap, most of the dust mass resides in the largest grains, since they are captured the most efficiently and therefore dominate the total dust mass. 
For these largest grains, $\Pi$ increases by a factor of $\sim230$, while for the smallest grains it grows by only about 35\%. 
Furthermore, the location of the peak of the dust-to-gas mass ratio depends on the grain size, varying from $\sim1.5R_\mathrm{p}$ (51 au) for the largest grains to $\sim1.3R_\mathrm{p}$ (44 au) for the smallest grains. 

The right panel of Figure \ref{fig:radial_D2G} shows the dust-to-gas mass ratio $\Pi$ of each species, but now plotted as a function of the height above the midplane.
These values are evaluated at the radial location where the total dust-to-gas mass ratio reaches its maximum ($\sim1.5 R_\mathrm{p}$).
As expected, the large dust grains settle towards the midplane with the smaller grains remaining extended at larger heights in the disk.
The vertical extent of each dust species has direct implications for whether it reaches the CPD.
This is discussed more in Section \ref{sec:res:CPD_dust}.

\subsection{Circumplanetary Disk Morphology and Kinematics}
\label{sec:res:CPD}

\begin{figure*}
\centering
    \includegraphics[width=0.8\textwidth]{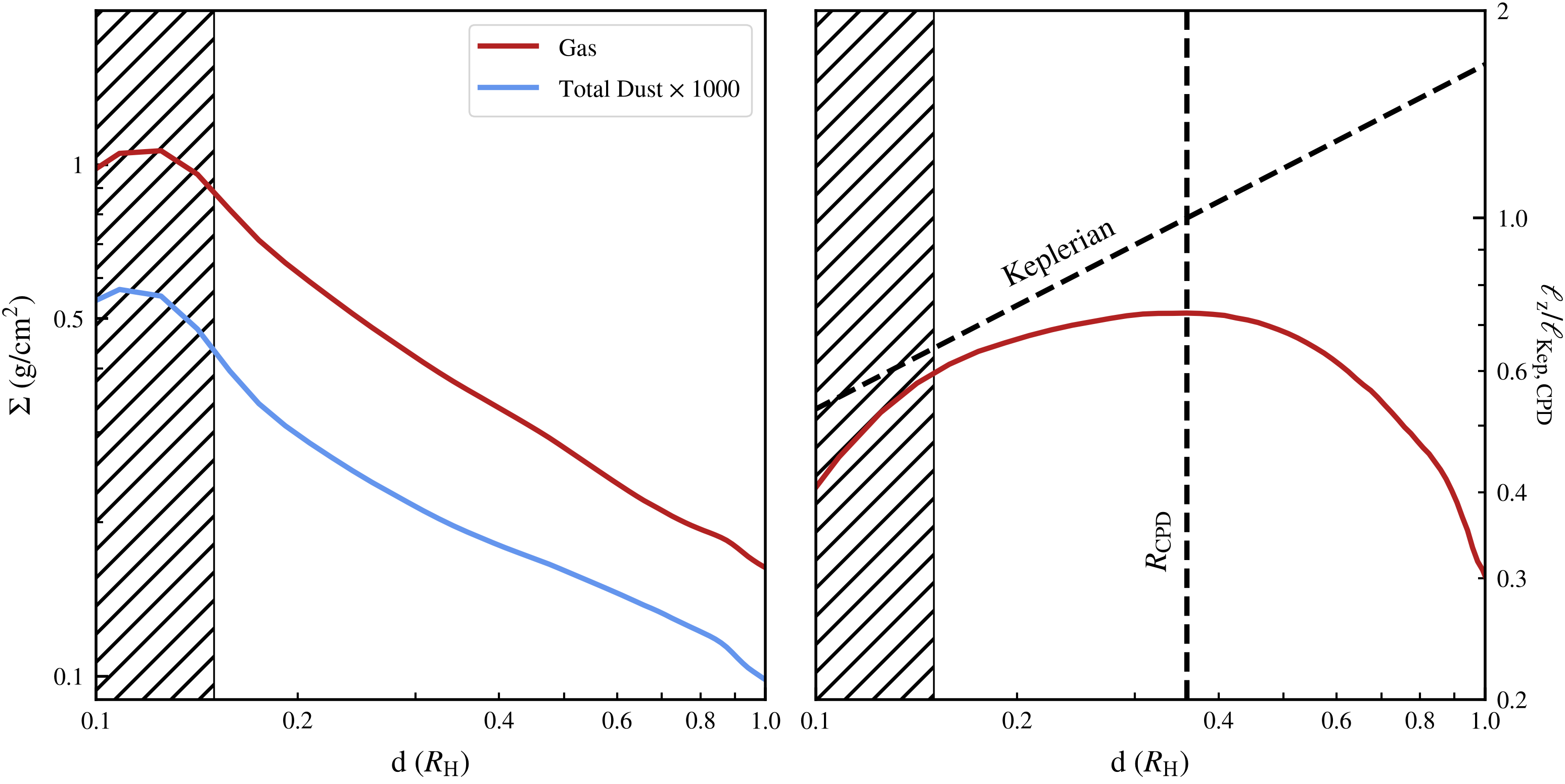}
    \caption{{\it Left:} Azimuthally averaged surface density of gas (red) and total dust (blue) within $R_\mathrm{H}$ for the 1 $M_\mathrm{J}$ model. The dust density is multiplied by 1000 to match the gas on the same scale. {\it Right:} Azimuthally averaged specific angular momentum of the gas at the disk midplane. The angular momentum is normalized to that of a Keplerian orbit around a 1 $M_\mathrm{J}$ planet at the CPD radius. This radius is calculated as the point where the gas angular momentum turns over ($\sim0.36R_\mathrm{H}$). Thus, the $R_{\mathrm{CPD}}$ line and the Keplerian line intersect at an $\ell_\mathrm{z}/\ell_{\mathrm{Kep,CPD}}$ value of 1. Within this radius, the angular momentum remains sub-Keplerian, indicating gas pressure support. In both panels, the black hatched region covers area interior to the sink radius.}
    \label{fig:CPD}
\end{figure*}

\begin{figure*}
    \centering
    \includegraphics[width=1.0\textwidth]{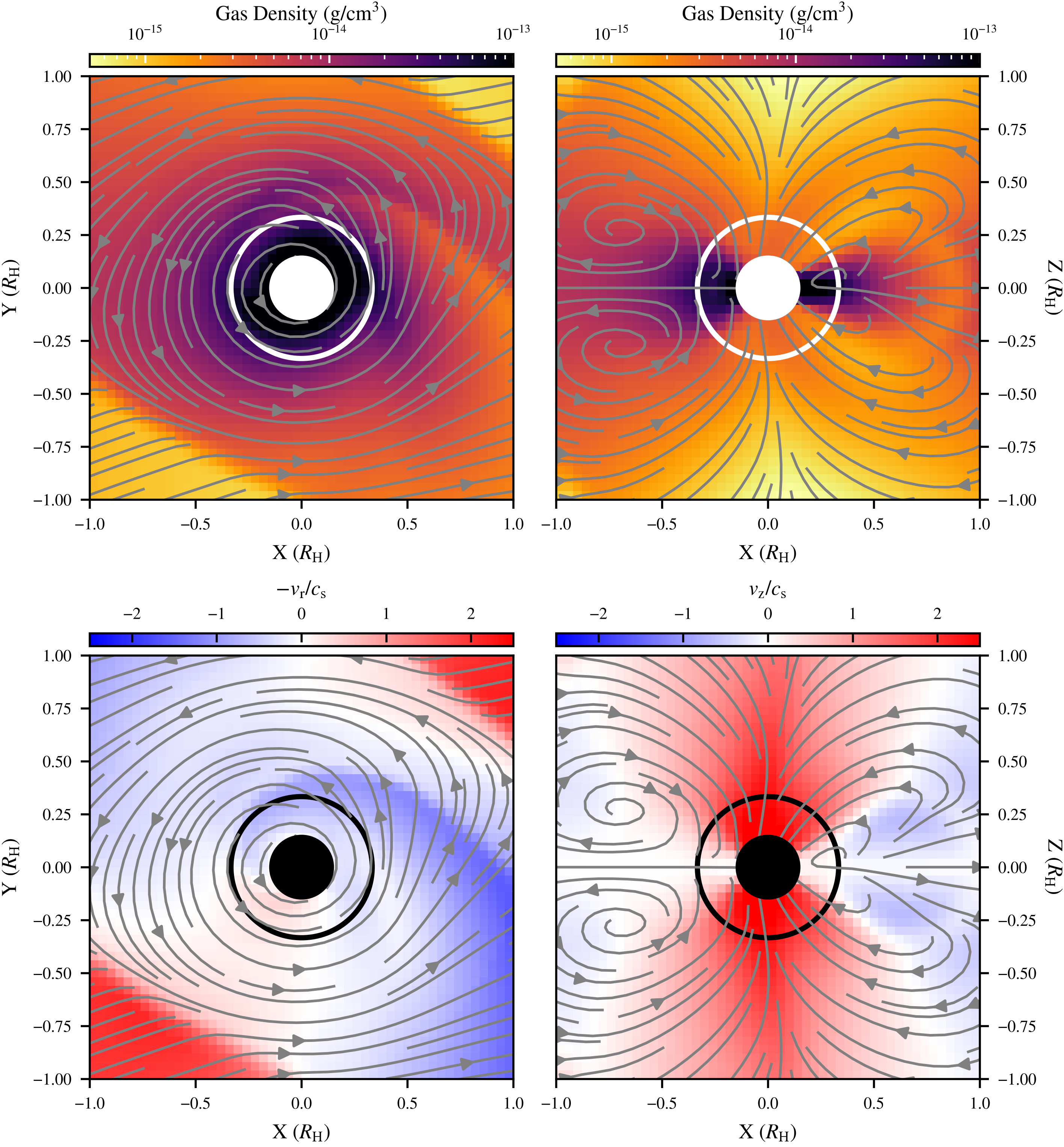}
    \caption{{\it Top:} Gas density in the midplane (left) and in the vertical slice (right) within 1 $R_\mathrm{H}$ of the planet. {\it Bottom Left:} Planet centered radial gas velocity in the midplane (color), we plot -$v_\mathrm{r}$ so that positive values corresponds to flow directed toward the planet, in accordance with our definition of $\dot{M}$. {\it Bottom Right:} Vertical gas velocity (color), where positive $v_\mathrm{z}$ indicates motion toward the midplane. The gray lines depict velocity streamlines projected onto the XY and XZ planes. The filled circle marks the sink radius, and the black ring has a radius of $1/3$ $R_\mathrm{H}$.}
    \label{fig:streamlines}
\end{figure*}

Figure \ref{fig:CPD} shows the azimuthally averaged gas and dust surface densities, along with the gas specific angular momentum in the disk midplane, evaluated within 1 $R_\mathrm{H}$ of the planet for the 1 $M_\mathrm{J}$ model. 
Beyond the sink region, denoted by the black hatched area, both gas and dust surface densities decrease approximately as $R^{-1}$, while the total dust-to-gas mass ratio in the CPD remains nearly constant at $\Pi_{CPD} \sim 0.05\%$. 
This extremely low value of $\Pi$ results directly from the dust trapping described in the previous section, which permits only the smallest grains to penetrate into the CPD. 
The total gas and dust masses contained within 1/3 $R_\mathrm{H}$, excluding the sink region, are $1.1\times10^{-4}$ and $5.2\times10^{-8}$ $M_\mathrm{J}$, respectively.
Following previous work, we define the outer radius of the CPD as the radius at which the specific angular momentum of the gas turns over \citep{Ayliffe+2009, Dangelo2003}. 
For the 1 $M_\mathrm{J}$ model, the turnover occurs at $ R=0.36 R_\mathrm{H}$, justifying our assumption of $ 1/3 R_\mathrm{H}$ as the indicative CPD outer radius throughout our analysis.
Furthermore, the comparison with the angular momentum of a perfectly Keplerian disk shows that the gas pressure plays a significant role in supporting the CPD. 
Both the surface density and angular momentum radial profiles show no abrupt discontinuities at the outer radius of the CPD. 
Instead, the simulation shows a gradual transition from the planet-dominated to the stellar-dominated regions. 

Although the azimuthally averaged radial profiles appear relatively simple, the full 3D density and kinematic structure of the material inside the Hill sphere is highly complex.
Figure~\ref{fig:streamlines} shows slices of the midplane and vertical gas density distribution within 1 $R_{\mathrm{H}}$ of the planet as well as the planet centered radial velocity and the vertical velocity directed toward the midplane.
We show negative planet centered $v_\mathrm{r}$ so that positive values indicate flow towards the planet, in accordance with our definition of $\dot{M}$.
Figure~\ref{fig:streamlines} has been time averaged over two full orbits of the planet with outputs every 0.01 orbits to ensure that any orbital phase dependent artifacts are smoothed out.
In the midplane (left column of Figure \ref{fig:streamlines}), the velocity streamlines show that material remains gravitationally bound and orbits the planet out to about $1/3$ $R_\mathrm{H}$. Beyond this radius, the flow instead streams past the CPD. 
In contrast to a standard CSD, the distributions of both gas and dust are markedly asymmetric in radius and azimuth.
The gas density in the midplane is characterized by at least one prominent spiral arm reaching the CPD outer radius.  

The midplane gas radial velocity as a function of azimuth changes between both inward and outward flow. 
Notably, outward-moving gas is concentrated along the spiral arm discussed above.
Conversely, most of the gas accreting on the CPD enters the Hill sphere from two opposite locations characterized by inward radial velocities between 0.5 and 1 times the local sound speed (bottom-left and top-right corners of the midplane $v_\mathrm{r}$ plot). 

The vertical slice through the Hill sphere (right column of Figure~\ref{fig:streamlines}) exhibits a meridional circulation pattern: gas flows radially inward in the upper layers of the disk and radially outward close to the midplane. 
This behavior agrees with previous studies \citep[e.g.,][]{Fung+2016, Szulagyi2022, Choksi+2023, Karlin2023} and is likewise observed in the dust velocity (see left column of Figure \ref{fig:combined}).
The vertical velocity displays a coherent structure: material near the planetary poles plunges toward the planet super sonically, while gas near the midplane is lifted back to higher altitudes.
The maximum vertical velocity achieved by the gas is nearly 3 times the sound speed.
This is in accordance with the results of \citet{Szulagyi+2016}, who find supersonic vertical flows near the poles that shock close to the planet and CPD surface. 
This is opposed to equatorial flows, which shock farther out at $\sim1-1.5$ $R_{\mathrm{H}}$ (see the standing shock in the bottom right panel of Figure \ref{fig:grid}).
However, our simulation resolution does not allow us to capture the shock of the vertically infalling material, which occurs well within 0.1 $R_{\mathrm{H}}$.

If we azimuthally average the density profile in the top right panel of Figure~\ref{fig:streamlines} and fit a vertical Guassian at $1/3 R_{\mathrm{H}}$, we find that the gas settles into a rather thick disk characterized by an aspect ratio ($H/R$) of $\sim0.36$. 
The aspect ratio at $1/3$ $R_\mathrm{H}$ calculated from the fixed temperature profile in Equation \ref{eq:planet_temp} is 0.43, indicating that the CPD is not quite in hydrostatic equilibrium.
These values are consistent with previous work and indicates that CPDs are puffier than CSDs \citep{Szulagyi+2014}.
However, it is worth noting that the vertical structure of the CPD depends on the assumed EOS, the local density, and the opacity.
As we use an isothermal equation of state, our CPD is indeed a disk.
However, studies adopting a different EOS (e.g., adiabatic) suggest that the CPD may instead be an even more puffed-up envelope \citep{Szulagyi+2016, Szulagyi2022, Krapp+2024}.


\subsection{Dust Accretion onto the Circumplanetary Disk}
\label{sec:res:CPD_dust}

\begin{figure*}[!t]
    \centering 
    \includegraphics[width=1.0\textwidth]{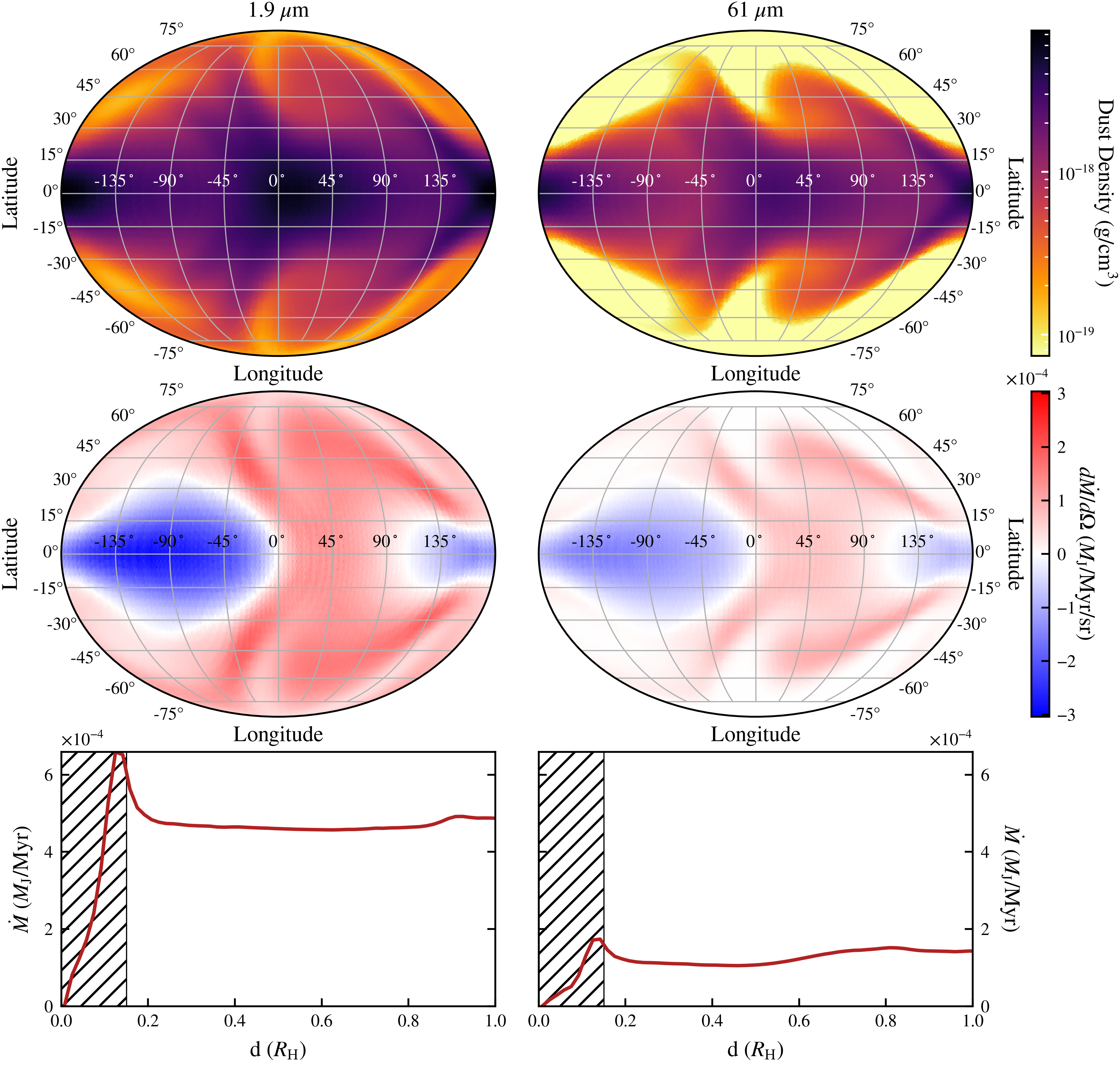}
    \caption{A collection of quantites within $R_\mathrm{H}$ for the 1.9 $\mu$m and 61 $\mu$m dust grains. {\it{Top:}} Mollweide projections of the dust density at 1/3 $R_\mathrm{H}$. 0$^\circ$ longitude points away from the star while $\pm180^\circ$ points towards it. Longitude increases counterclockwise around the planet (see the left column of Figure \ref{fig:streamlines} for reference). {\it{Middle:}} The same as the top, but now for the mass accretion rate per unit solid angle of each species in the frame of the planet. Positive (red) values indicate flow directed toward the planet.
    {\it{Bottom:}} The vertically and azimuthally integrated mass accretion rate of each species as a function of the distance from the planet. The black hatched region covers area interior to the sink radius.}
    \label{fig:Money}
\end{figure*}

\begin{deluxetable}{ccc}[!t]
\tablehead{
\colhead{Species} &
\colhead{$\dot{M}(1\,M_\mathrm{J})$} &
\colhead{$\dot{M}(2.5\,M_\mathrm{J})$}
}
\startdata
Gas & 3.4 & 3.1 \\
Total Dust & 1.9$\times10^{-3}$ & 1.0$\times10^{-3}$ \\
0.61 $\mu$m & -- & 2.7$\times10^{-4}$\\
1.9 $\mu$m & 4.7$\times10^{-4}$ & 3.8$\times10^{-4}$ \\
4.8 $\mu$m & -- & 2.5$\times10^{-4}$\\
6.1 $\mu$m & 6.7$\times10^{-4}$ & --  \\
10 $\mu$m & -- & 1.2$\times10^{-4}$ \\
19 $\mu$m & 6.0$\times10^{-4}$ & --  \\
22 $\mu$m & -- & 9.5$\times10^{-6}$ \\
61 $\mu$m & 1.2$\times10^{-4}$ & 3.0$\times10^{-8}$ \\
190 $\mu$m & 1.1$\times10^{-6}$ & 2.1$\times10^{-8}$  \\
610 $\mu$m & 2.1$\times10^{-8}$ & 1.8$\times10^{-8}$  \\
\enddata
\caption{Mass accretion rates onto the $1/3$ $R_\mathrm{H}$ sphere in $M_\mathrm{J}$/Myr calculated by the amount of mass being removed by the sink. The different columns correspond to different simulations. \label{tab:mdot}}
\end{deluxetable}

\begin{figure*}[!t]
    \centering 
    \includegraphics[width=1.0\textwidth]{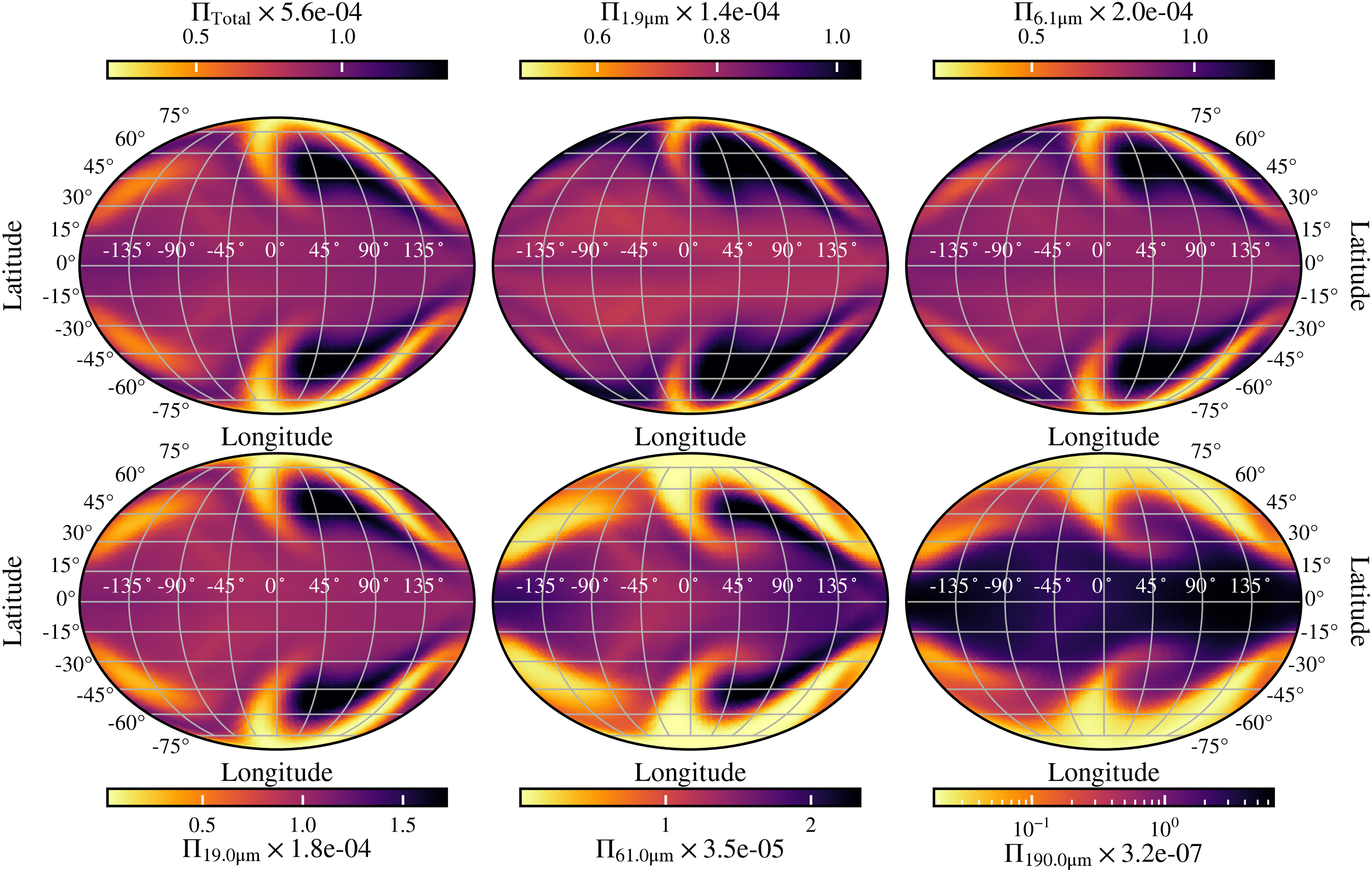}
    \caption{Mollweide projections of the time-averaged dust-to-gas mass ratios over the surface of a sphere with radius equal to $1/3$ $R_\mathrm{H}$. Longitudes of $\pm180^\circ$ indicate the direction pointing toward the central star, while 0$^\circ$ corresponds to the direction pointing away from the star. Longitude increases as you move counterclockwise around the planet. Each species' dust-to-gas mass ratio is normalized to the dust-to-gas mass ratio of the incoming material from Table \ref{tab:mdot}.}
    \label{fig:Mollweide}
\end{figure*}

To study how dust accretes onto the CPD, we first analyze the spatial distribution of the dust density and mass accretion rates at 1/3 $R_\mathrm{H}$.
Figure \ref{fig:Money} shows Mollweide projections of the dust density (top row) and mass accretion rates per unit solid angle (middle row) for the 1.9 $\mu$m and 61 $\mu$m dust grains. 
The full $\theta$ and $\phi$ integrated mass accretion rate is also shown (bottom row).

Comparing the dust densities in the 1.9 $\mu$m and 61 $\mu$m projections underscores the joint effects of dust settling and filtration. 
Even though the 61 $\mu$m grains initially contain about $\sim6$ times more mass, their density at $1/3\,R_H$ is lower than that of the 1.9 $\mu$m grains. 
This decrease is evident throughout the full $1/3\,R_\mathrm{H}$ sphere, with the 61 $\mu$m dust density particularly reduced near the poles, consistent with the behavior of a larger grain that settles more strongly.

These larger grains enter the CPD at this lower altitude, as shown in the middle row of Figure \ref{fig:Money}. 
There is essentially no high-altitude inflow of large grains, while the small grains can accrete across all altitudes, though the accretion rate is suppressed near the poles. 
The mass accretion rate reaches its minimum along “whiskers” that extend from the poles toward lower latitudes. 
In the dust density projections, these whiskers appear as regions of pronounced dust absence.

Both grain sizes accumulate material near the midplane between longitudes of $0^\circ$ and $90^\circ$. 
This quadrant corresponds to the red inflow region seen in the bottom left panel of Figure \ref{fig:streamlines} at the 1/3 $R_\mathrm{H}$ circle.
Material captured over this longitude interval is supplied directly from the outer disk and delivered onto the CPD. 
Nonetheless, most of the dust enters the CPD via high-latitude hotspots. 
For the 1.9 $\mu$m grain, these hot spots occur at longitudes of roughly $-45^\circ$ to $0^\circ$ and $45^\circ$ to $135^\circ$, spanning latitudes from about $75^\circ$ down to $15^\circ$. 
The 61 $\mu$m grain exhibits comparable structures, though they are shifted toward lower latitudes and are characterized by lower mass accretion rates. 
These regions constitute the main channels by which dust is accreted onto the CPD. 
In addition, both dust populations exhibit a strong outflow (blue regions) concentrated near the midplane, between longitudes $135^\circ$ and $-45^\circ$. 
This corresponds to dust exiting the CPD along the leading spiral arm as it wraps around the planet (bottom left panel of Figure \ref{fig:streamlines}).

Although material both enters and leaves the CPD, the bottom row of Figure \ref{fig:Money} indicates that the mass accretion rates integrated over the full sphere remain positive and roughly steady across $R_\mathrm{H}$.  
This suggests that the measurements taken at 1/3 $R_\mathrm{H}$ are comparable to the mass extracted by the sink.  
The 1/3 $R_\mathrm{H}$ accretion rates for the gas and for each dust species in both simulations are reported in Table \ref{tab:mdot}.  
To more accurately resolve the abrupt boundary between grains that traverse the gap and grains that are trapped in the outer disk, the 2.5 $M_\mathrm{J}$ run adopts extra, closely spaced dust bins at the small-size end.  
Because dust is trapped in the circumstellar disk, the CPD accretes only a minor portion of the total dust reservoir. 
For the 1 $M_\mathrm{J}$ case, the dust delivered to the CPD is dominated by grains smaller than 61$\mu$m, summing to approximately $1.9 \times 10^{-3}$ $M_\mathrm{J}$/Myr. 
In the 2.5 $M_\mathrm{J}$ case, the accreted dust is primarily even finer ($<10$ $\mu$m) and the total dust accretion rate decreases to $1\times 10^{-3}$ $M_\mathrm{J}$/Myr.

In order to compare gas and dust CPD structures, we calculate the dust-to-gas mass ratio of the material that enters the circumplanetary region at $1/3$ $R_\mathrm{H}$ and again utilize a Mollweide projection for display (Figure \ref{fig:Mollweide}).
Like Figures \ref{fig:streamlines} and \ref{fig:Money}, Figure \ref{fig:Mollweide} has been time averaged over two full orbits of the planet.
Overall, the dust-to-gas mass ratio is extremely low ($\sim 5\times 10^{-4}$) and varies with both longitude and latitude.
As discussed in Appendix~\ref{app:2.5mj}, the overall dust-to-gas mass ratio for the 2.5 $M_\mathrm{J}$ model is even lower, reaching $\sim 2\times 10^{-4}$.
The total dust-to-gas mass ratio (top-left panel) is lowest toward the CPD polar regions and along the whiskers seen in Figure \ref{fig:Money} that extend to lower latitudes. 
Importantly, the maximum dust-to-gas mass ratio is achieved at high latitudes in the hotspots from Figure \ref{fig:Money} between 45\arcdeg \ and 60\arcdeg \ and longitudes between 45\arcdeg \ and 90\arcdeg \, indicating that most of the dust accreted by the CPD does not enter through the disk midplane.   

The dust-to-gas mass ratio varies with the grain size, with smaller grains showing more uniform values across the $1/3$ $R_\mathrm{H}$ spherical surface. 
Small grains are more strongly coupled to the gas and therefore more vertically extended, allowing for more uniform flow at higher latitudes.
As the grain size increases, the dust-depleted polar region extends to lower latitudes, and the spatial variations become more pronounced.
Eventually, for the largest grains, the dust-to-gas mass ratio is highest along the disk midplane, as these larger dust grains are more settled.
However, it should be noted that the amount of dust supplied to the CPD in the form of particles larger than $\sim50$ $\mu$m is negligible compared to that of smaller particles. 
 
To quantify the size-dependent dust filtration responsible for the CPD's solid depletion seen in Figures \ref{fig:Money} and \ref{fig:Mollweide}, we utilize the mass accretion rates calculated in Table \ref{tab:mdot}. 
Following previous studies \citep[e.g.,][]{Weber2018}, we define a filtration efficiency as:

\begin{equation}
    \text{FE} = 1 - \frac{\dot{M}_\mathrm{d,CPD}/\dot{M}_\mathrm{g,CPD}}{\dot{M}_\mathrm{d,CSD}/\dot{M}_\mathrm{g,CSD}}.
\end{equation}
Where $\dot{M}_\mathrm{d,CPD}$ and $\dot{M}_\mathrm{g,CPD} $ are the dust and gas accretion rates at 1/3 $R_\mathrm{H}$ respectively.
$\dot{M}_\mathrm{d,CSD}$ and $\dot{M}_\mathrm{g,CSD}$ are the dust and gas accretion rates in the outer circumstellar disk.
By defining the filtration efficiency in this way, we directly compare the rate material is flowing through, or being trapped in, the outer disk to the rate at which it is removed by the planet.
A filtration efficiency of 0 means the CPD accretes dust and gas at the exact relative rates supplied by the outer disk, whereas an efficiency of 1 means that virtually no dust is accreted when compared to the gas. 

\begin{figure}[!t]
    \centering
    \includegraphics[width=1.0\linewidth]{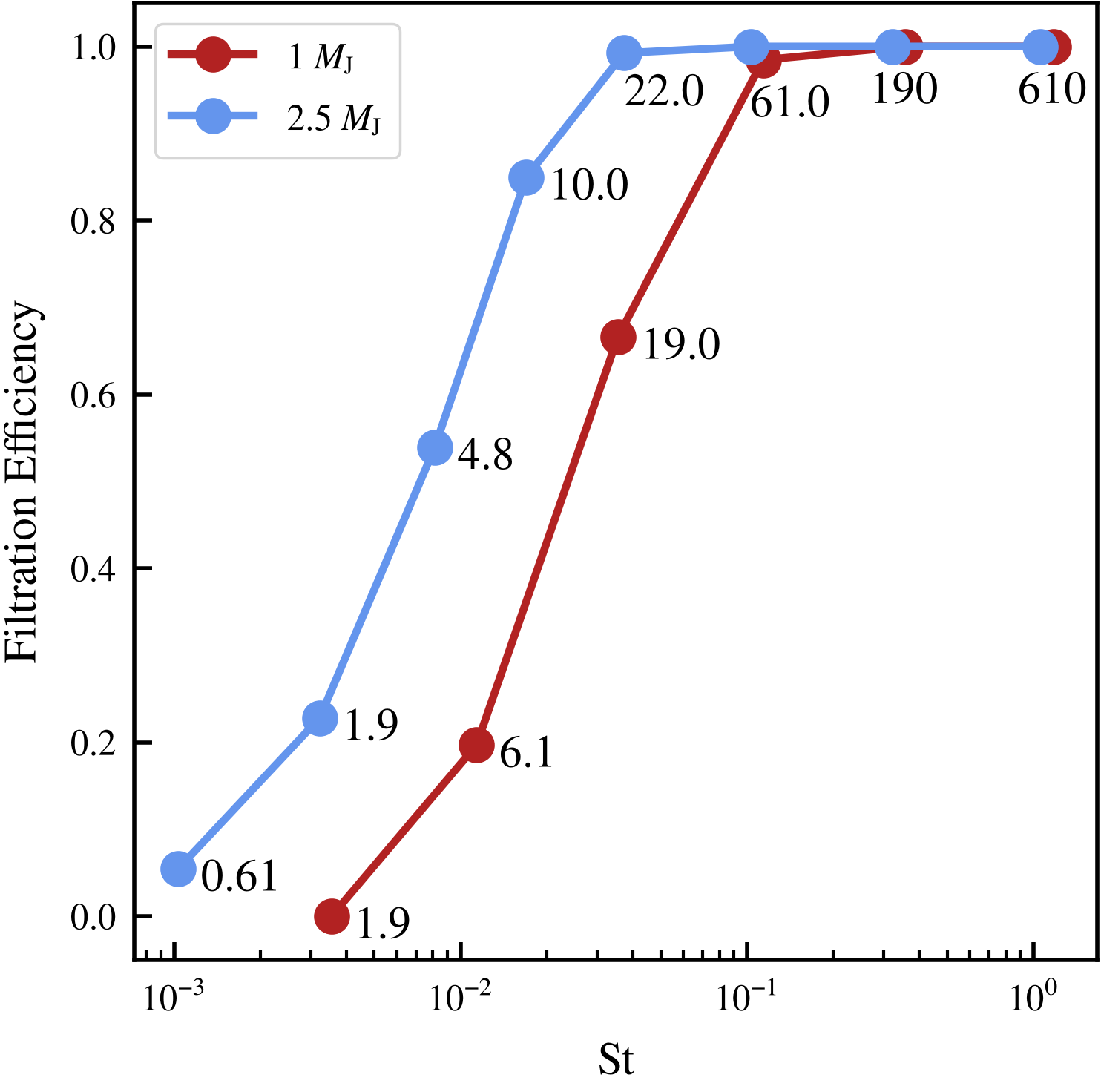}
    \caption{Filtration efficiency for dust species with different sizes in the 1 $M_\mathrm{J}$ and 2.5 $M_\mathrm{J}$ models as a function of the Stokes number (St) calculated from the gas surface density at the center of the dust trap. The dust trap is at 1.5 $R_\mathrm{p}$ and 1.75 $R_\mathrm{p}$ for the 1 $M_\mathrm{J}$ and 2.5 $M_\mathrm{J}$ simulations, respectively. 
    The dust sizes in micron are also marked on the plot.}
    \label{fig:Filtration}
\end{figure}

As shown in Figure~\ref{fig:Filtration}, the filtration efficiency increases with the grain size (and thus Stokes number), eventually capping at full filtration for grains larger than 61~$\mu$m and 22~$\mu$m, for the 1 $M_\mathrm{J}$ and 2.5 $M_\mathrm{J}$ models, respectively. 
At the gas surface densities encountered in the center of the dust trap at $R=1.5 R_\mathrm{p}$, these grains have Stokes parameters between 0.1 and 0.03, respectively.
However, even smaller grains are subject to significant depletion. 
While 1.9 $\mu$m grains fully diffuse through the gap in the 1 $M_\mathrm{J}$ case, they experience a filtration efficiency of about 0.25 in the $2.5M_\mathrm{J}$ model. 
Taken at face value, these results suggest that CPDs around more massive planets not only accrete less dust overall, but the accreted dust is also composed of much smaller grains. 
The implications of these results for moon formation and ALMA observations are discussed in the next section. 

\section{Discussion}
\label{sec:discussion}

The dust accretion rates and filtration efficiencies discussed above provide critical insights into how the circumplanetary disk around PDS~70~c may acquire solid material, directly constraining the feasibility and timescales of rocky moon formation. 
Furthermore, this severe, size-dependent dust filtration fundamentally shapes the disk's observational signature in millimeter-wave continuum emission because it determines the disk opacity.

\subsection{Implications for Moon Formation around PDS~70~c}
\label{sec:disc:moon_form}

In the literature regarding the formation of Jupiter's Galilean moons, the gas-drag assisted capture and subsequent ablation of planetesimals dynamically scattered from the circumstellar disk into the circumplanetary disk is often considered the standard mechanism for solid enrichment \citep[e.g., ][]{Estrada+2009, Mosqueira+2010, Fujita+2013, Dangelo+2015, Ronnet+2020}.
For example, numerical integrations by \cite{Ronnet+2018} demonstrated that planetesimals captured within a Jovian circumplanetary disk are subject to ablation due to frictional heating with the gas \citep[depending on local gas density and temperature, see][]{Dangelo+2015}, thereby supplying the disk with small dust grains.

To understand why this capture mechanism may be problematic for wide-separation systems like PDS 70 c, we must examine the specific conditions under which it operates efficiently.
The simulations by \citet{Ronnet+2020} were tailored to a Jupiter analog at an orbital distance of 5.2 au, where the circumplanetary disk is relatively compact and dense. 
In their setup, the gas surface density at the CPD outer radius was assumed to be $2 \times 10^2\,\mathrm{g~cm}^{-2}$. 
However, the outer radius of a circumplanetary disk scales directly with the planetary Hill radius, which is proportional to the planet's orbital distance. 
For a given circumplanetary disk-to-planet mass ratio, the gas surface density within the disk scales inversely with the square of its outer radius ($\Sigma_{\mathrm{out}} \propto R_{\mathrm{out}}^{-2}$). 
Because PDS 70 c orbits much further out at 34 au, its Hill radius is significantly larger than that of Jupiter, resulting in a substantially more diffuse circumplanetary disk. 
Consequently, our models yield a gas density about 400 times lower than that of typical Jovian models. 
Since the gas drag required to capture planetesimals is directly proportional to the local gas density, this process is inherently much less efficient at these wide separations.

Furthermore, an underlying assumption of the capture model is the availability of a background reservoir of planetesimals (from which the planet itself started to form). 
Recent planetesimal population synthesis models by \citet{Lenz+2019} parameterized planetesimal formation efficiency as a function of the radial flux of drifting pebbles. 
Their results demonstrate that planetesimal formation can be highly efficient in the inner disk, but drops off precipitously at larger radial distances because the spatial spacing of pebble traps and the pebble flux itself depend strongly on orbital separation. 
Consequently, {\it in situ} planetesimal formation at wide orbits (e.g., beyond 30 au) is expected to be inherently inefficient, meaning the requisite population of large background planetesimals might not exist in sufficient numbers around PDS~70~c. 
Therefore, the continuous accretion of small, filtered dust grains directly from the circumstellar disk, as modeled in our simulations, could represent an important solid-delivery mechanism for forming satellite systems around PDS 70 c, and wide-separation planets in general.
In the absence of any dust growth, the formation of moons, however, would still require the presence of initial ``seed'' bodies in the CPD onto which small solids can accrete.

The dust accretion rates and filtration efficiencies derived in Section~\ref{sec:res:CPD_dust} directly constrain how circumplanetary disks acquire this solid material. To evaluate whether the PDS~70~c circumplanetary disk can form a satellite system comparable to Jupiter's Galilean moons (which contain approximately $2 \times 10^{-4} M_\mathrm{J}$ in solids), we must examine the instantaneous mass contained within the disk. 
In our 1 $M_\mathrm{J}$ simulation, the gas surface density outside the sink sphere ($R \ge 0.15 R_\mathrm{H}$) follows a power-law profile of roughly $R^{-1}$. 
Measuring directly from 0.15 $R_\mathrm{H}$ to the disk's outer edge at 1/3 $R_\mathrm{H}$, the enclosed gas and dust masses are $1.1 \times 10^{-4} M_\mathrm{J}$ and $5.2 \times 10^{-8} M_\mathrm{J}$, respectively, corresponding to an extremely low circumplanetary dust-to-gas mass ratio of approximately 0.05\%. 

Because the orbits of the Galilean satellites lie at roughly 0.008 to 0.035 (Jupiter) $R_\mathrm{H}$, well inside our unresolved sink sphere, we must extrapolate this $R^{-1}$ profile inward to estimate the mass of the moon-forming region.
Integrating this extrapolated profile yields an additional $0.9 \times 10^{-4} M_\mathrm{J}$ of gas, bringing the total circumplanetary gas mass to $2.0 \times 10^{-4} M_\mathrm{J}$.
Assuming the dust-to-gas mass ratio remains a uniform 0.05\%, the total instantaneous dust mass in the extrapolated disk is only about $1.0 \times 10^{-7} M_\mathrm{J}$.
This instantaneous solid reservoir is more than three orders of magnitude smaller than the mass required to build a Galilean-like satellite system.
We note that this extrapolation assumes the temperature structure (and thus pressure support) measured in the resolved disk continues at smaller radii. 
In reality, planetary thermal emission and accretion heating likely increase the disk temperature near the planet, which could modify the pressure balance and alter the equilibrium surface density profile from simple $R^{-1}$ scaling. 
However, even if the inner disk temperature were a factor of several times higher than our resolved measurements, the resulting change in mass would be modest compared to the three orders of magnitude discrepancy between our instantaneous solid reservoir and the mass required to build a Galilean-like satellite system.

For a constant mass accretion rate, the instantaneous gas and dust mass within the circumplanetary disk depends heavily on the assumed viscosity, at least in the absence of internal dust traps. 
Our hydrodynamic models adopt a constant kinematic viscosity corresponding to an $\alpha$ of 0.01 at the planet's location. 
If the actual viscosity in the circumplanetary region is lower, the surface density would correspondingly increase, allowing the disk to retain a higher steady-state mass. 
However, even with a moderately lower viscosity, the instantaneous mass is likely insufficient to form moons on its own, reinforcing the necessity of a continuous inflow of material coupled with an efficient internal mechanism to halt radial drift. 

Modern models of satellite formation broadly agree that in situ accretion of the Galilean moons at their current close-in orbital distances is highly unlikely \citep[e.g.,][]{Canup+2006, Batygin+2020}. 
Instead, these moons likely formed at a much greater orbital radius and subsequently migrated inward. 
For instance, \citet{Drazkowska+2018} modeled dust evolution and satellitesimal formation in a circumplanetary disk and found that an outward gas flow creates a natural pressure trap at a distance of 85 $R_\mathrm{J}$ (approximately 0.11 $R_\mathrm{H}$) from the planet. 
In their models, this outward gas flow counteracts radial drift long enough for infalling dust to accumulate, triggering the streaming instability and continuously converting the incoming dust into satellitesimals right at this 85 $R_\mathrm{J}$ mark.

The dust supply rate assumed in these previous models, however, differs significantly from the filtered accretion rates we calculate since the hydro simulations on which they are calibrated lack dust. 
Instead, \citet{Drazkowska+2018} assumed an unfiltered gas infall rate of 2.0 $M_\mathrm{J}$/Myr with a standard dust-to-gas mass ratio of 0.01, yielding an incoming dust supply rate of $2.0 \times 10^{-2} M_\mathrm{J}$/Myr. 
By contrast, our 3D multifluid model explicitly captures the gap-edge pressure maximum, which filters the accreting flow and reduces the solid accretion rate onto the 1 $M_\mathrm{J}$ planet's disk to $1.9 \times 10^{-3} M_\mathrm{J}$/Myr. 
While our filtration-reduced rate is an order of magnitude lower, it is still sufficient to deliver the $2 \times 10^{-4} M_\mathrm{J}$ Galilean moon inventory in roughly $10^5$ years, provided that an internal trap successfully converts this steady influx of small grains into larger bodies.
In this case, however, the formation mechanism could not rely on pebble accretion, which for small dust has an efficiency $< 10$\% \citep[e.g.,][]{Dangelo+2024}.
Unfortunately, our global 3D models cannot directly resolve the internal dust evolution needed to test this accumulation, because the trapping region proposed by \citet{Drazkowska+2018} at 0.11 $R_\mathrm{H}$ lies entirely within the artificial sink sphere, a region not directly modeled. 
Furthermore, a more complex treatment of radiation would be required to model the effects of planetary and viscous heating. 

The outlook for moon formation changes dramatically for the 2.5 $M_\mathrm{J}$ planet model. 
Our simulations demonstrate that more massive planets carve deeper gaps with stronger pressure gradients, making dust filtration far more severe. 
For the 2.5 $M_\mathrm{J}$ case, the total dust accretion rate drops to $1.0 \times 10^{-3} M_\mathrm{J}$/Myr and is restricted almost entirely to grains smaller than 10 $\mu$m. 
Because more massive planets effectively starve their own circumplanetary disks of solid material, accumulating the necessary mass for moon formation takes substantially longer. 
This suggests a potential anti-correlation between a giant planet's mass and the presence or mass of rocky moons it can host, as the most massive planets efficiently cut off their own solid supply lines. 

\subsection{Scaling our results}

\begin{deluxetable*}{cccc}[!t]
\tablehead{
\colhead{Fluid Species} &
\colhead{Rescaled Disk Mass} &
\colhead{Rescaled q = -2.5} &
\colhead{Rescaled $a_{\mathrm{max}} = 1\mathrm{mm}$}
}
\startdata
Gas & 0.34 & 3.4 & 3.4 \\
Total Dust & 5.9$\times10^{-5}$ & 7.5$\times10^{-6}$ & 5.9$\times10^{-3}$\\
0.19 $\mu$m &  1.5$\times10^{-5}$ & -- & -- \\
0.61 $\mu$m &  2.1$\times10^{-5}$ & -- & -- \\
1.9 $\mu$m  & 1.9$\times10^{-5}$ & 2.8$\times10^{-7}$ & 1.5$\times10^{-3}$\\
6.1 $\mu$m & 3.8$\times10^{-6}$ & 1.3$\times10^{-6}$ & 2.1$\times10^{-3}$\\
19 $\mu$m & 3.5$\times10^{-8}$ & 3.6$\times10^{-6}$ & 1.9$\times10^{-3}$\\
61 $\mu$m & 6.6$\times10^{-10}$ & 2.2$\times10^{-6}$ & 3.8$\times10^{-4}$\\
190 $\mu$m & -- & 6.5$\times10^{-8}$ & 3.5$\times10^{-6}$\\
610 $\mu$m & -- & 3.9$\times10^{-9}$ & 6.7$\times10^{-8}$\\
\enddata
\caption{Rescaled accretion rates from our 1 $M_\mathrm{J}$ simulation in $M_\mathrm{J}/$Myr. For the Rescaled mass column, we take the 1 $M_\mathrm{J}$ simulation and reduce the disk mass by a factor of 10. This means that a grain now behaves as if its size were 1/10 of its original size. There is an additional factor of $\sqrt{10}$ due to the smaller dust grain taking up a smaller portion of the total dust size distribution by mass. Both the q=-2.5 and $a_{\mathrm{max}} = 1\mathrm{mm}$ columns serve to redistribute the dust mass into different bins, thereby changing the ammount that gets filtered versus accreted. \label{tab:rescaled_mdot}}
\end{deluxetable*}

Crucially, the satellite formation timescale is highly sensitive to the assumed dust population properties, such as the Stokes parameter ($\mathrm{St}$), the grain size distribution slope ($q$), and the maximum grain size of the population ($a_{\mathrm{max}}$). 
To study the effect of these parameters, we rescale our $1\,M_\mathrm{J}$ simulation in three distinct ways: adjusting the disk mass and density, altering $q$, and changing $a_{\mathrm{max}}$. 
Table~\ref{tab:rescaled_mdot} summarizes the accretion rates resulting from these rescalings.  

Because our simulations employ a locally isothermal equation of state and exclude self-gravity, the underlying hydrodynamics scale directly with the assumed disk mass. 
Scaling the gas mass down by a factor of 10 increases $\mathrm{St}$ for a given physical grain size, strongly altering the dust dynamics. 
As demonstrated in Table~\ref{tab:rescaled_mdot}, a lower disk mass leads to a heavily reduced solid accretion rate. 
For example, a $19\,\mu\mathrm{m}$ grain now behaves the same as a $190\,\mu\mathrm{m}$ grain did previously, as both share the same $\mathrm{St}$. 
Looking at Figure~\ref{fig:Filtration}, the $190\,\mu\mathrm{m}$ grain had $\mathrm{St} \sim 0.35$ and was completely filtered out of the CPD region. 
This is now the fate of the rescaled $19\,\mu\mathrm{m}$ grain, which was previously only partially filtered. Thus, the surviving accreted grains are much smaller, with the planet accumulating appreciable amounts of dust only for sizes smaller than $10\,\mu\mathrm{m}$ after rescaling. 
The effect of this rescaling is two-fold: the disk itself is less massive (meaning less dust to begin with), and even more of the larger, mass-carrying grains are trapped in the circumstellar disk. 
The rescaling also allows one to estimate the time-dependent delivery of dust to the CPD region, as the surrounding disk dissipates.

While our fiducial model assumes a standard MRN size distribution slope of $q = -3.5$, observations suggest that dust in protoplanetary disks often grows to produce shallower slopes closer to $q = -2.5$ \citep{Dalessio+2001}. 
If we rescale our fiducial $1\,M_\mathrm{J}$ accretion rates using a size distribution slope of $q = -2.5$, the solid accretion rate drops precipitously. 
Because a shallower slope places a larger fraction of the total solid mass into larger grains, which are efficiently filtered by the gap-edge pressure maximum, very little mass ultimately reaches the planet. 
Under these conditions, accumulating enough solids to build massive moons or match the PDS 70 c dust mass would require over $10\mathrm{~Myr}$, exceeding typical protoplanetary disk lifespans.  

Similarly, a more evolved dust population will possess a larger $a_{\mathrm{max}}$ as grains coalesce. 
Models of multiwavelength PDS 70 observations suggest that the maximum grain size of the dust population is $\sim 1\mathrm{~cm}$ \citep{Sierra+2025}, motivating our choice in this study. 
However, given the uncertainties inherent in opacity modeling, we rescale our simulation to evaluate the effect of a smaller maximum grain size of $a_{\mathrm{max}} = 1\mathrm{~mm}$. 
This increases our dust accretion rates by a factor of $\sim 3$, showing that dust is accreted even more rapidly when more mass is contained within grains that are not filtered at the gap edge.  

These results suggest that a massive disk with a relatively unevolved dust population (more akin to Class 0/I systems) can supply solids to a CPD much more readily. 
On one hand, if the planet formed very early on, the CPD region could be too warm for volatile-carrying grains to survive.
On the other, if satellite formation occurs later, during the Class II phase, the CPD must possess efficient mechanisms to retain the small dust it accretes. 

\subsection{Implications for ALMA Observations}

This size-dependent filtration has direct consequences for the observational detectability of circumplanetary disks. 
Because massive giant planets act as strict filters, the dust that successfully populates their circumplanetary disks consists exclusively of small particles. 
At submillimeter wavelengths, such as ALMA Band 7 (855 $\mu$m), these small grains possess absorption opacities much smaller than millimeter-sized pebbles.
For example, the DSHARP size-averaged absorption opacity at 855$\mu$m drops by a factor of $\sim4.5$ when $a_{\mathrm{max}}$ changes from 1mm to 10$\mu$m.
In their analysis, \citet{Benisty+2021} estimate a range of PDS 70 c dust masses that corresponds to these opacity differences.

Consequently, deriving circumplanetary dust masses from submillimeter continuum fluxes under the assumption of a canonical, large-grain population will severely underestimate the true dust mass present in the disk. 
Interpreting millimeter continuum detections around massive protoplanets, therefore, requires invoking significantly larger total dust masses to account for the low opacity of the surviving, heavily filtered small grains. 
By extension, these findings must be taken into account when estimating upper limits on CPD masses from ALMA non-detections \citep[e.g., the WISPIT 2 b case;][]{Facchini+2026} or when calculating the required sensitivity to detect a CPD. 
Because more massive planets impose stricter filtration, their CPDs will be intrinsically much fainter at millimeter wavelengths for a given total dust mass. 
Our results, therefore, suggest that future searches for circumplanetary disks at millimeter wavelengths should perhaps focus on lower-mass planets, which allow larger, higher-opacity grains to cross the gap and replenish the disk.  

Furthermore, this severe dust depletion fundamentally alters how we infer total disk mass from continuum observations. 
Standard observational estimates assume a primordial dust-to-gas mass ratio of 1\% ($\Pi = 0.01$). 
However, our 2.5 $M_\mathrm{J}$ model indicates that the CPD dust-to-gas mass ratio is heavily depleted, reaching only $\sim 0.03\%$ ($\Pi = 3 \times 10^{-4}$). 
If our estimates of the dust-to-gas mass ratio in the outer CPD are correct, extrapolating total disk mass from dust continuum data using the standard 1\% factor will drastically underestimate the amount of gas present.  
For example, ALMA observations of PDS 70 c infer a CPD dust mass between $(2-9) \times 10^{-5}\,M_\mathrm{J}$, depending on the assumed grain size \citep{Benisty+2021}. 
If we assume the dust in this CPD is dominated by small grains and apply the dust-to-gas mass ratio from our 2.5 $M_\mathrm{J}$ model, the inferred gas mass of the PDS~70~c CPD increases to $0.06 - 0.3\,M_\mathrm{J}$. 
This implies that the CPD could contain a significant fraction of the central planet's own mass.

Although the PDS~70~c CPD has not been detected in molecular line emission, $^{13}$CO emission has been detected from a candidate CPD in AS 209 \citep{Bae+2022}.
These observations suggest that the candidate CPD might have a mass larger than a few to tens of percent of the planet's mass. 
Furthermore, because no continuum is detected toward the candidate CPD, the authors claim that the dust-to-gas mass ratio in the CPD is very small, $\sim10^{-4}$ for a 1 $\mu$m maximum grain size.
While more evidence is needed to confirm the AS 209 candidate, these results align with our simulations and represent one physical scenario for the CPD's gas and dust content: a depleted-dust CPD with a large gas mass.
If CPDs are indeed so massive, ALMA molecular line observations could therefore provide a powerful tool for detecting and characterizing them. 

Another plausible scenario is that dust is efficiently trapped within the CPD. 
If internal traps efficiently accumulate the incoming small dust grains over long timescales without accumulating gas at the same rate, the effective steady-state dust-to-gas mass ratio within the CPD would rise significantly above the $0.03\%$ value we see in our simulations. 
This scenario may apply to PDS~70~c, where there is a clear detection of the dust continuum, but no detection of molecular line emission yet. 
In this case, both high-resolution CPD simulations that test CPD dust trapping theories, like those suggested in \citet{Drazkowska+2018}, and high-resolution deep molecular line images that spatially separate a gas CPD from the circumstellar disk, are needed before any definitive conclusion can be made.

Finally, our proposed anti-correlation between CPD dust mass (and maximum grain size) and the mass of the central planet provides a clear theoretical prediction.
This anti-correlation could, in principle, be directly tested once a sufficiently large sample of young giant planets is observed at high resolution.



\subsection{Caveats and Limitations}
\label{sec:disc:caveats}

While our high-resolution 3D hydrodynamic simulations provide robust constraints on the size-dependent filtration of dust at the gap edge, our physical framework relies on certain approximations that warrant consideration. 
Because constraints related to the grid resolution (i.e., the artificial sink sphere) and the assumed dust population properties are discussed in the previous sections, we focus here on the unaddressed thermodynamic and hydrodynamic physics.

First, our simulations employ a locally isothermal equation of state, which does not capture complex thermodynamic effects self-consistently. 
Studies incorporating adiabatic or fully radiative equations of state have demonstrated that planetary luminosity and accretion shocks can significantly alter the local thermal structure, potentially resulting in a much more vertically extended, "puffed up" circumplanetary disk, more similar to an envelope \citep{Szulagyi+2016, Szulagyi2022, Krapp+2024}.
A hotter, thicker disk could alter the meridional circulation patterns and impact the disk's ability to retain accreting solids.  

Second, we only utilize one level of $\nu$ for all of our simulations.
At lower values of $\nu$, the pressure gradient at the gap edge would be stronger and would more efficiently trap dust grains.
The overall accretion rate throughout the disk would also decrease.
Thus, a lower viscosity could drop the measured planetary accretion rates precipitously.
Because our gas accretion rates onto PDS 70 c are on the higher end of observed values, we are in a high viscosity limit and our reported values should be regarded as upper limits.

Finally, our analysis assumes the circumplanetary disk is fed exclusively by the continuous radial drift and diffusion of small dust grains. 
We neglect alternative solid enrichment pathways, such as the direct capture and ablation of large planetesimals ($a > 100\text{ m}$) dynamically scattered by the outer disk \citep{Zhou+2007, Shiraishi+2008, Tanigawa+2014, Dangelo+2015, Suetsugu+2016, Suetsugu+2017, Ronnet+2018}.  
However, because in-situ planetesimal formation at wide orbits (e.g., $> 34\text{ au}$) is expected to be inefficient, the accretion of small dust grains may still represent the dominant mass-delivery mechanism for forming satellite systems.

\section{Conclusions}

We performed 3D hydrodynamic simulations to quantify the mass accretion rates of gas and dust onto the circumplanetary disk of PDS~70~c. 
By combining observationally constrained global disk parameters with high-resolution AMR grids focused within the planetary Hill sphere, we tracked the size-dependent filtration efficiencies of solids crossing the circumstellar gap. 

Our main conclusions are as follows:  
\begin{enumerate}
\item Size-dependent filtration: Dust accretion onto the CPD is highly fractionated by grain size. 
The gas pressure maximum at the outer edge of the planetary gap efficiently traps large grains, resulting in a circumplanetary dust-to-gas mass ratio that is roughly two orders of magnitude lower than the initial circumstellar ratio ($\Pi = 0.02$).  

\item Influence of planet mass: The filtration efficiency is highly sensitive to the planetary mass. 
While the 1 $M_\mathrm{J}$ model allows grains up to $\sim61 \mu$m to accrete efficiently, the stronger pressure trap in the 2.5 $M_\mathrm{J}$ model restricts the accreted solid mass almost entirely to grains smaller than $10 \mu$m. 
Even perfectly coupled $1.9 \mu$m grains in the 1 $M_\mathrm{J}$ simulation show signs of filtration around the 2.5 $M_\mathrm{J}$ planet.  



\item Reservoir of solids for satellite formation: Despite the severe filtration of large grains, the mass accretion rates are sufficient to deliver the estimated solid mass of the PDS~70~c CPD ($\sim10^{-4} M_\mathrm{J}$) over a period of $\sim 10^{5}$ years.
This mass would also suffice to produce a satellite system similar to the Galilean moons, provided all those solids can be converted into satellites.

\item Observational implications: Because massive giant planets effectively choke off the supply of large grains, the resulting CPDs are heavily populated by small grains that possess low opacities at submillimeter wavelengths. Therefore, matching observed ALMA millimeter fluxes in systems like PDS~70~c may require significantly more total dust mass than models assuming well-mixed, large-grain populations would suggest.  

\end{enumerate}
A.I. and C.H.G. acknowledge support from the National Aeronautics and Space Administration under grant No. 80NSSC18K0828.
C.H.G. acknowledges support from the Rice Space Institute Center for Planetary Origins to Habitability (RSI-CPO2H) Graduate Fellowship program.
C.H.G., H.L., S.L., G.D., and A.M.D acknowledge the support from LANL/LDRD.
This work was supported by the U.S. Department of Energy through the Los Alamos National Laboratory.
Los Alamos National Laboratory is operated by Triad National Security, LLC, for the National Nuclear Security Administration of U.S. Department of Energy (Contract No. 89233218CNA000001).
This research used resources provided by the Los Alamos National Laboratory Institutional Computing Program, which is supported by the U.S. Department of Energy National Nuclear Security Administration under Contract No. 89233218CNA000001.
Research presented in this article was supported by the Laboratory Directed Research and Development program of Los Alamos National Laboratory under project number 20240039DR.
We acknowledge the use of LLMs, Claude and ChaGPT, for minor refinements to the languauge of this paper.

\appendix

\section{2D Hydrodynamic Simulations}

In this appendix, we detail the procedure and results of 2D simulations that were run to calibrate our eventual model of the PDS 70 disk.

\subsection{2D Simulation Setup}
\label{sec:app:2D_meth}

To calibrate our 3D models, we performed 2D locally isothermal simulations using the LA-COMPASS hydrodynamic code, which solves multi-fluid equations including the coupling between gas and dust \citep{Li+2005, Li+2009}. 
To ensure consistency with our subsequent 3D simulations, we adopt a constant kinematic viscosity, $\nu$. 
Because the \citet{SS+1973} $\alpha$-disk model is inherently 2D, adopting a constant $\nu$ disk is more appropriate for our goals of extending to 3D simulations. 
Furthermore, we adopt a constant initial gas surface density, $\Sigma_{\mathrm{g},0}$, for our 2D simulations for the same reasons outlined in Section \ref{sec:3D_Disk}. 

To calculate a realistic disk temperature profile for our simulations, we employ the 3D radiative transfer code RADMC3D \citep{Dullemond+2012} to compute the disk thermal structure and corresponding sound speed.   
RADMC3D uses dust densities and opacities to perform a thermal Monte Carlo simulation, sampling photon packets from a stellar blackbody distribution to calculate the dust equilibrium temperature. 
We calculate the disk temperature iteratively as follows. 
First, we calculate the dust density as:
\begin{equation}
\label{eq:dust_rho}
\rho_{\mathrm{d}}(R,z) = \frac{\Pi_d\Sigma_{\mathrm{g},0}(R)}{\sqrt{2\pi}H_\mathrm{g}(R)} e^{-z^2/2H_\mathrm{g}(R)^2},
\end{equation}
assuming the gas pressure scale height $H_\mathrm{g}(R) = 1 \textrm{au} \left(\frac{R}{20\textrm{au}}\right)^{5/4}$.
Here the gas and dust are assumed to be well mixed.

Second, we derive the corresponding dust temperature using RADMC3D and DSHARP size averaged dust opacities with a maximum grain size of 1~cm and a grain size distribution $n(a) \propto a^{-3.5}$ \citep{Birnstiel+2018}. 
We then calculate the vertically integrated, density-weighted radial temperature profile and gas sound speed \citep[see, e.g., equation A3 in][]{Bae2019} and a new disk pressure scale height $H_\mathrm{g}(R) = c_\mathrm{s}(R)/\Omega(R)$. 
Finally, the updated $H_\mathrm{g}(R)$ is used to recalculate the dust density via Eq.~\ref{eq:dust_rho} and, using RADMC3D, a new temperature profile. We repeat this iteration a few times until the disk temperature varies by less than 1\%.

For the initial velocities, we prescribe the gas radial velocity according to the self-similarity solution of a constant viscosity, constant surface density viscous disk \citep{Lynden+1974}:
\begin{equation}
    v_{\mathrm{R,g}} = -\frac{3}{2}\frac{\nu}{R}.
\end{equation} 
The initial gas azimuthal velocity is calculated taking into account the effect of the gas pressure as:
\begin{equation}
    \frac{v_{\mathrm{\phi,g}}^2}{R} = \frac{GM_\star}{R^2} + \frac{1}{\rho}\frac{\partial P}{\partial R}.
    \label{eq:az_vel}
\end{equation}
The initial dust radial velocity follows the prescription of  \citet{Birnstiel+2010}:
\begin{equation}
    v_{\mathrm{R,d}} = \frac{v_{\mathrm{R,g}}}{1+\text{St}^2} + \frac{v_{\mathrm{d}}}{\text{St} + \text{St}^{-1}}, 
\end{equation}
where $v_\mathrm{d}$ denotes the drift velocity, defined as:
\begin{equation}
    v_\mathrm{d} = \frac{\frac{\partial P}{\partial R}}{\Sigma_\mathrm{g} \Omega_\mathrm{k}}
\end{equation}
Finally, the initial azimuthal velocity of the dust is 
\begin{equation}
    v_{\mathrm{\phi,d}} = v_\mathrm{k} + \frac{\left(v_{\mathrm{\phi,g}} - v_\mathrm{k}\right)}{1+\text{St}^2}
\end{equation}
so that grains with $\textrm{St}\gg 1$ rotate around the star at Keplerian velocity while grains with $\textrm{St}\ll 1$ are coupled to the gas. 

The 2D hydrodynamic simulations include 9 fluids: the gas and 8 dust species, with sizes ranging from 1~$\mu$m to 1~cm, spaced by 0.5 dex. 
Dust grains smaller than 1~$\mu$m are assumed to be coupled to the gas.
The dust is initialized everywhere with the same mass per bin prescription as our 3D simulations which is outlined in Appendix \ref{sec:app:dust}.

Using the prescribed initial disk density and velocity, and the RADMC3D temperature profile, we carry out 2D locally isothermal simulations that include two embedded planets representing PDS~70~b and PDS~70~c.
Following \citet{Bae2019}, we fix the mass of PDS~70~b to 5 $M_\mathrm{J}$, while PDS~70~c is assigned a mass of either 1 or 2.5 $M_\mathrm{J}$.
The planets are located at 22 and 34 au, respectively, and their orbits are kept fixed.
This configuration is motivated by \citet{Bae2019}, who showed that such an orbital arrangement is dynamically stable.
We use a typical gravitational softening length of $0.6H$ \citep{Muller2012}.
For the 2D computational domain, we adopt 672 logarithmically spaced radial cells spanning 6–198 au and 936 cells to resolve the full 2$\pi$ azimuthal range.
Outflow boundary conditions are applied in the radial direction, while azimuthal boundaries are taken to be periodic.
Both the 1 $M_\mathrm{J}$ and 2.5 $M_\mathrm{J}$ models are evolved for 22,300 orbits at 20 au, corresponding to a total simulation time of $\sim2$ Myr.

After running our 2D, multifluid hydrodynamical simulations and evolving the density of multiple dust species, we derive the 3D dust density for each species as follows.
We take the evolved 2D dust surface density, $\Sigma_\mathrm{d}(R,\phi,a)$, and use Eq.~\ref{eq:dust_rho} to calculate $\rho_\mathrm{d}(R, \phi, z, a)$. 
In this step, the gas surface density is replaced with the dust surface density, and the dust-to-gas mass ratio is subsequently dropped. 
The gas pressure scale height is replaced with the dust pressure scale height, which accounts for vertical settling and is given by \citep{Youdin2007}:
\begin{equation}
H_\mathrm{d}(R, a) = H(R) \left( 1+ \frac{\textrm{St}(R,a)}{\alpha(R)}\right)^{-1/2},
\end{equation}
where:
\begin{equation}
    \alpha(R) = \frac{\nu}{c_\mathrm{s}(R)H(R)},
\end{equation}
is the \citet{SS+1973} $\alpha$ viscosity.
Here it varies as a function of $R$ because we utilize a constant kinematic viscosity, $\nu$.

Finally, adopting DSHARP single-grain opacities for each dust species in our 2D simulation, we run 3D radiative transfer calculations with RADMC3D to generate synthetic continuum emission at the wavelength of the ALMA Band 7 observations \citep[855 $\mu$m, ][]{Benisty+2021}.



\subsection{2D Observational Constraints on PDS 70}
\label{sec:app:2D_res}

\begin{figure}[!t]
    \centering
    \includegraphics[width=1.0\columnwidth]{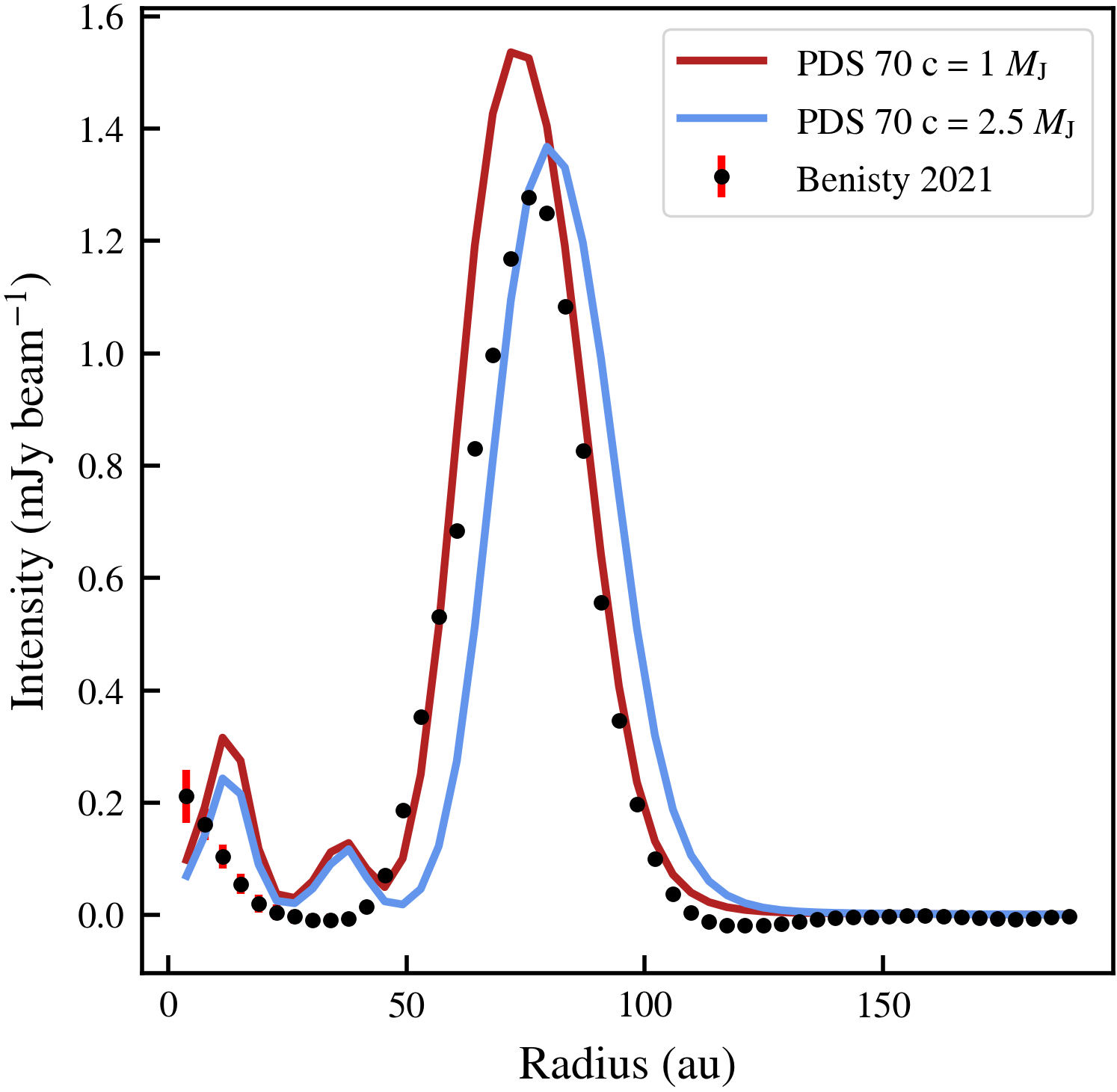}
    \caption{Azimuthally averaged radial intensity profiles of the 855 $\mu$m dust continuum emission recorded toward PDS~70 \citep{Benisty+2021} compared with our 2D hydrodynamic models. 
    }
    \label{fig:av_prof}
\end{figure}

To compare our synthetic models with existing observations, we evaluate their image's Fourier transform at the same {\it uv}-coordinates as the ALMA data from \citet{Benisty+2021}, using the CASA task \texttt{ft}.
We subsequently reconstruct a corresponding model image with \texttt{tclean}. 
This procedure guarantees that the spatial sampling characteristics of the ALMA dataset are accurately reproduced.
Both the observational and model datasets are imaged with Briggs weighting and a robust parameter of 2.0, yielding a synthesized beam with an FWHM of 0.071\arcsec$\times$0.062\arcsec. 
We then deproject the images using an inclination of 51.7\arcdeg\ and a position angle of 160.4\arcdeg\ \citep{Keppler2019}, and compute azimuthally averaged profiles to obtain the radial intensity distributions shown in Figure~\ref{fig:av_prof}. 

The best agreement between the simulated and observed radial profiles is obtained for an initial gas surface density $\Sigma_\mathrm{g} = 1.2$ g cm$^{-2}$, a dust-to-gas mass ratio $\Pi=0.02$, and a kinematic viscosity $\nu = 1.2\times10^{16}$ cm$^2$ s$^{-1}$. 
These values correspond to a Shakura–Sunyaev parameter of $\alpha = 10^{-2}$ at the orbital radius of PDS 70 c and $\alpha = 5\times10^{-3}$ at the radius of the dust ring. 
The resulting initial mass accretion rate through the CSD is $\dot{M} \sim 2 \times 10^{-6}$ $M_\mathrm{J}$ yr$^{-1}$, broadly consistent with the accretion rates inferred by \citet{Choksi+2025}. 
The midplane temperature profile is $T(r)= 41.6 \textrm{K} \left( \frac{r}{20\textrm{au}} \right)^{-0.6}$ and the corresponding gas pressure scale height is $H(R) = 1.26 \textrm{ au} \left(\frac{R}{20\text{ au}}\right)^{1.2}$,
which implies an aspect ratio of 7\% at the location of PDS 70 c (34 au).
The corresponding midplane temperature profile is in good agreement with the thermal structure derived from molecular line observations by \citet{Law2024}.

As illustrated in Figure \ref{fig:av_prof}, the 1 $M_\mathrm{J}$ PDS 70 c model correctly reproduces the observed radial location of the continuum ring, but it overpredicts its peak intensity by about 15\%. 
In contrast, the 2.5 $M_\mathrm{J}$ model yields a closer agreement with the azimuthally averaged brightness profile, although the radius of the ring maximum is shifted to slightly larger values than observed. 
Both simulations match the observed ring width within the measurement uncertainties. 
Overall, these findings suggest that our adopted PDS 70 disk–planet configuration captures the primary characteristics of the ALMA continuum observations and serves as a suitable reference for the subsequent 3D simulations.

\section{Viscous boundary conditions}
\label{app:bound_cond}
In this appendix, we detail the implementation of our viscous boundary condition. 
At the inner radial boundary, we first set all of the components of the velocity in the ghost zones to their initial value.
We then calculate the mass accretion rate in the first active cell based on its current density, the initial viscous radial velocity, and its area.
We set the density in the ghost zone so that this accretion rate is constant given the initial radial velocity and area of the ghost cell.
We follow the same procedure for dust, instead using its initial radial velocity and density. 

At the outer radial boundary, we use the viscous, torque-free boundary from \citet{Dempsey+2020a} but extended to 3D, spherically-symmetric mass inflow. We first calculate the initial and current angular momentum and density in the last active cell.
We then calculate the initial angular momentum and density in the ghost cells.
We seek to inject a constant angular momentum flux at the outer boundary by setting the ghost cell density as:

\begin{equation}
    \rho_\mathrm{g} = \rho_\mathrm{a}\left(\frac{\rho_{\mathrm{g,0}} l_{\mathrm{a,0}}}{\rho_{\mathrm{a,0}}l_{\mathrm{g,0}}}\right) + \frac{\rho_{\mathrm{g, 0}}\left(l_{\mathrm{g,0}} - l_{\mathrm{a,0}}\right)}{l_{\mathrm{g,0}}}
\end{equation}

\noindent Where $\rho$ and $l$ are the densities and angular momenta.
The subscript $\mathrm{g}$ denotes ghost values while $\mathrm{a}$ denotes active.
The 0 subscript denotes an initial disk value.
The first term extrapolates and rescales the active density using the initial densities and angular momenta to ensure the angular momentum flux from the active domain smoothly transitions into the ghost cells.
The second term corrects this value to match the intended initial steady-state $\dot{M}$.
With the density set, we then set the radial velocity in the ghost zone by scaling it as:
\begin{equation}
    v_{r,\mathrm{g}}  = v_{\mathrm{r,g},0}\left(\frac{\rho_{\mathrm{g},0}}{\rho_\mathrm{g}}\right)
\end{equation}
The azimuthal and vertical velocities are set to their initial values.

We also implement wave-killing zones at the inner and outer radial boundaries to damp spiral density waves excited by the planet and prevent wave reflections.
We follow the procedure outlined in \citet{Dempsey+2020a} where only the radial velocity is damped to conserve angular momentum.
We damp the radial velocity to the initial conditions following the formula in the appendix of \citet{Dempsey+2020a}:
\begin{equation}
    \frac{\partial v_\mathrm{r}}{\partial t} = - \left(\frac{v_\mathrm{r} - v_{\mathrm{r},0}}{\tau_\mathrm{d}}\right)R(r)
\end{equation}
Here $\tau_{d}$ is the local damping timescale with $\tau_\mathrm{d} = 1 /(30\Omega_\mathrm{k}(r))$.
$R(r)$ is the quadratic damping function outlined in \citet{ValBorro+2006}.
We set the inner (outer) edge of the wave-killing zones to be 0.5 (3.5) $r_\mathrm{p}$.

\section{Dust Mass Distribution and Size Selection}
\label{sec:app:dust}

For the dust distribution utilized in our simulations, we assume an $a_{\mathrm{min}}$ of 0.1 $\mu$m and an $a_{\mathrm{max}}$ of 1 cm, motivated by \citet{Sierra+2025}.
We distribute the dust in size with an MRN slope of -3.5 \citep{MRN1977}.
Since we are interested in the size and mass of dust that makes it to the CPD region, we need a fine spacing in St to capture the transition from coupled to decoupled dust grains.
Because of this, we space the edges of our dust bins by 0.5 dex rather than the usual 1 dex (i.e. 1 $\mu$m to 3.16 $\mu$m, 3.16 $\mu$m to 10 $\mu$m, etc.).
For a grain size distribution with a slope of -3.5, the fraction of the total dust mass in each bin, $m_\mathrm{b}$, is:
\begin{equation}
    m_{\mathrm{b}} = \frac{\sqrt{a_\mathrm{u}} - \sqrt{a_\mathrm{l}}}{\sqrt{a_{\text{max}}} - \sqrt{a_{\text{min}}}}, 
\end{equation}
where $a_\mathrm{u}$ is the upper bound of the dust bin and $a_\mathrm{l}$ the lower.
We then calculate a representative size for each dust bin such that the mass of dust in the bin on either side of this representative size is the same.
For the MRN distribution this means:

\begin{equation}
    \sqrt{a_\mathrm{u}} - \sqrt{a_\mathrm{r}} = \sqrt{a_\mathrm{r}} - \sqrt{a_\mathrm{l}},
\end{equation}
where $a_\mathrm{r}$ is the representative size of the bin.
We then solve for this size, obtaining:

\begin{equation}
    a_\mathrm{r} = \left(\frac{\sqrt{a_\mathrm{u}} + \sqrt{a_\mathrm{l}}}{2}\right)^2
\end{equation}

\noindent Calculating the representative size for 10 bins from 0.1 $\mu$m to 1 cm yields 0.19 $\mu$m, 0.61 $\mu$m, 1.9 $\mu$m, 6.1 $\mu$m, etc.

For the 2D simulations, we disregard the grain sizes below 1$\mu$m, assuming that they are well coupled to the gas, and focus on the larger 8 sizes.
For our 3D simulations, results shown in Figure \ref{fig:Filtration} indicated that the 2.5 $M_\mathrm{J}$ simulation was initially undersampled in St.
Because of this, we split the two bins covering grains from $10^{-3.5}$cm to $10^{-2.5}$ cm into 3 bins, now spaced by 0.33 dex.
The size selection and mass distribution follows the same as before.
We also track grains of size $0.61\mu m$ in the 2.5 $M_\mathrm{J}$ simulation because even the $1.9 \mu m$ grains show signs of filtration.

\section{Results for the 2.5 $M_\mathrm{J}$ planet mass model}
\label{app:2.5mj}
In this section, we reproduce the main figures from the paper, but for the case of the 2.5 $M_\mathrm{J}$ planet.
Qualitatively, most of the figures remain the same, but there are slight changes due to the larger gravitational influence of the planet.
We also show how some of the quantitative measurements change as a result of this.

\begin{figure}[!t]
    \centering
    \includegraphics[width=1.0\textwidth]{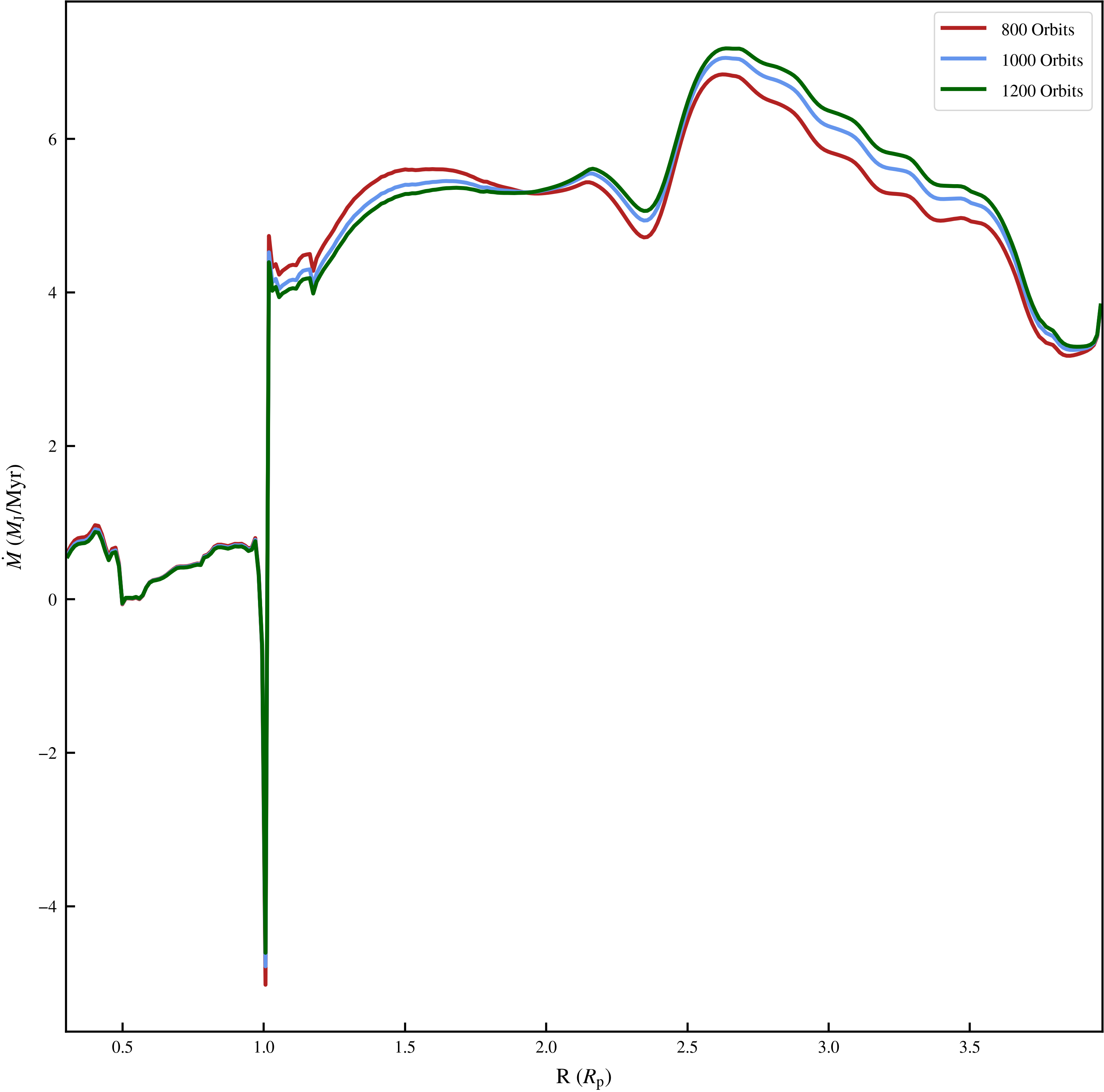}
    \caption{The same as Figure \ref{fig:global_mdot}, but now for the case of the 2.5 $M_\mathrm{J}$ planet.
    }
    \label{fig:global_mdot_2p5}
\end{figure}

In Figure \ref{fig:global_mdot_2p5}, we again show the global accretion profile throughout the disk.
Like the 1 $M_\mathrm{J}$ case, there is little variation in the global accretion rate throughout the disk after $\sim1000$ orbits.
The $\dot{M}$ being supplied to the disk at the outer boundary is slightly higher in this simulation than the 1 $M_\mathrm{J}$ simulation.
This is because the more massive planet deposits more angular momentum into the disk.
This angular momentum is carried outward until it is deposited by our damping zone in the outer disk, increasing the local angular momentum there and causing more material to move outward.
The increased density in the outer disk is picked up by our viscous boundary condition which then increases the density of the ghost zones. 

\begin{figure*}[!]
    \centering
    \begin{subfigure}[b]{1.0\textwidth}
        \centering
        \includegraphics[width=0.9\textwidth]{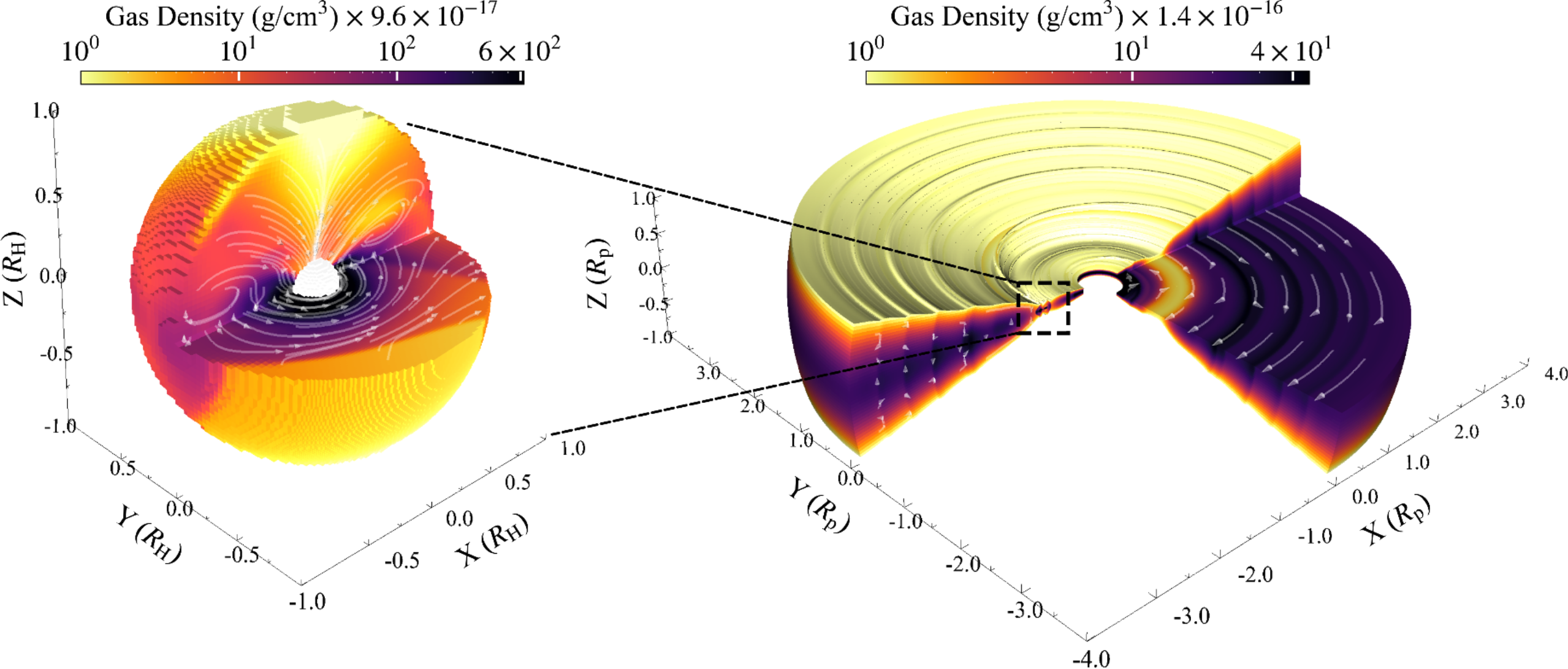}
        \label{fig:sub1_2p5}
    \end{subfigure}
    \par\vfill
    \begin{subfigure}[b]{1.0\textwidth}
        \centering
        \includegraphics[width=0.9\textwidth]{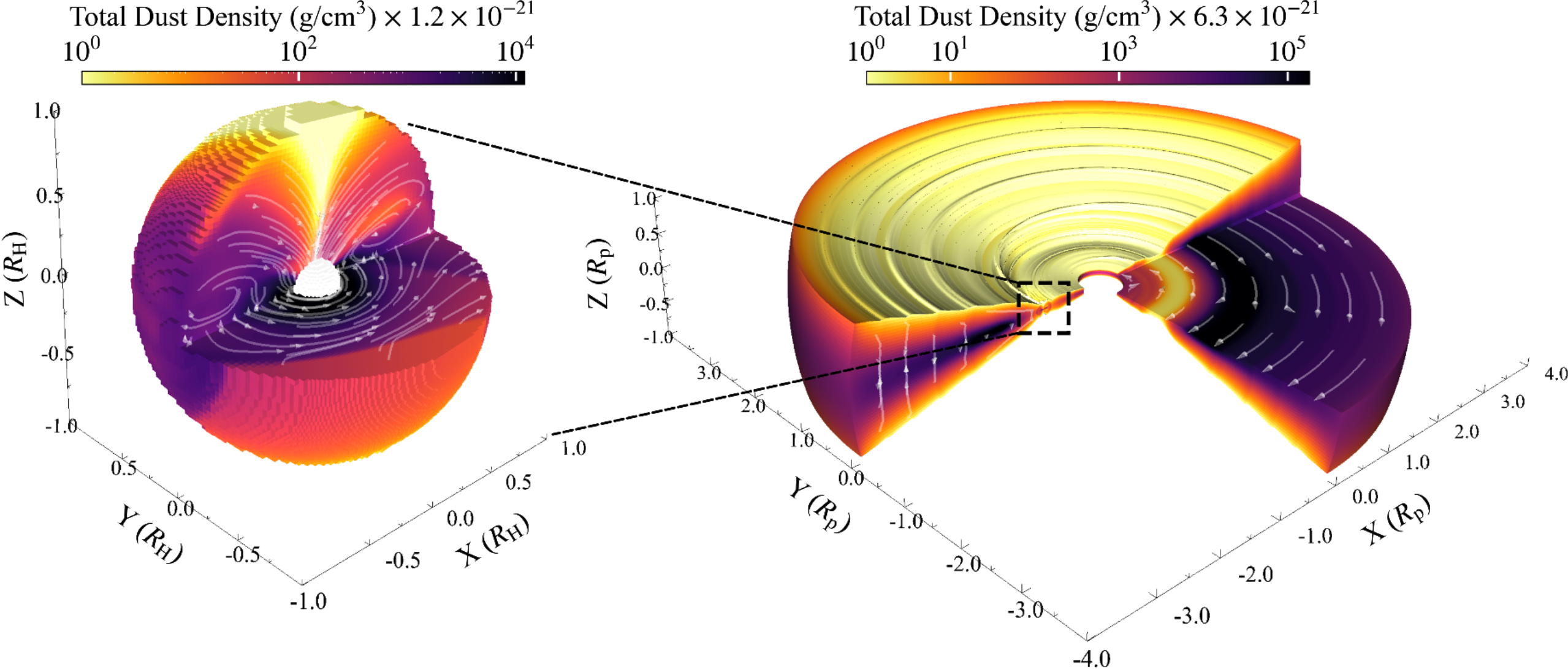}
        \label{fig:sub2_2p5}
    \end{subfigure}
    \par\vfill
    \begin{subfigure}[b]{1.0\textwidth}
        \centering
        \includegraphics[width=0.9\textwidth]{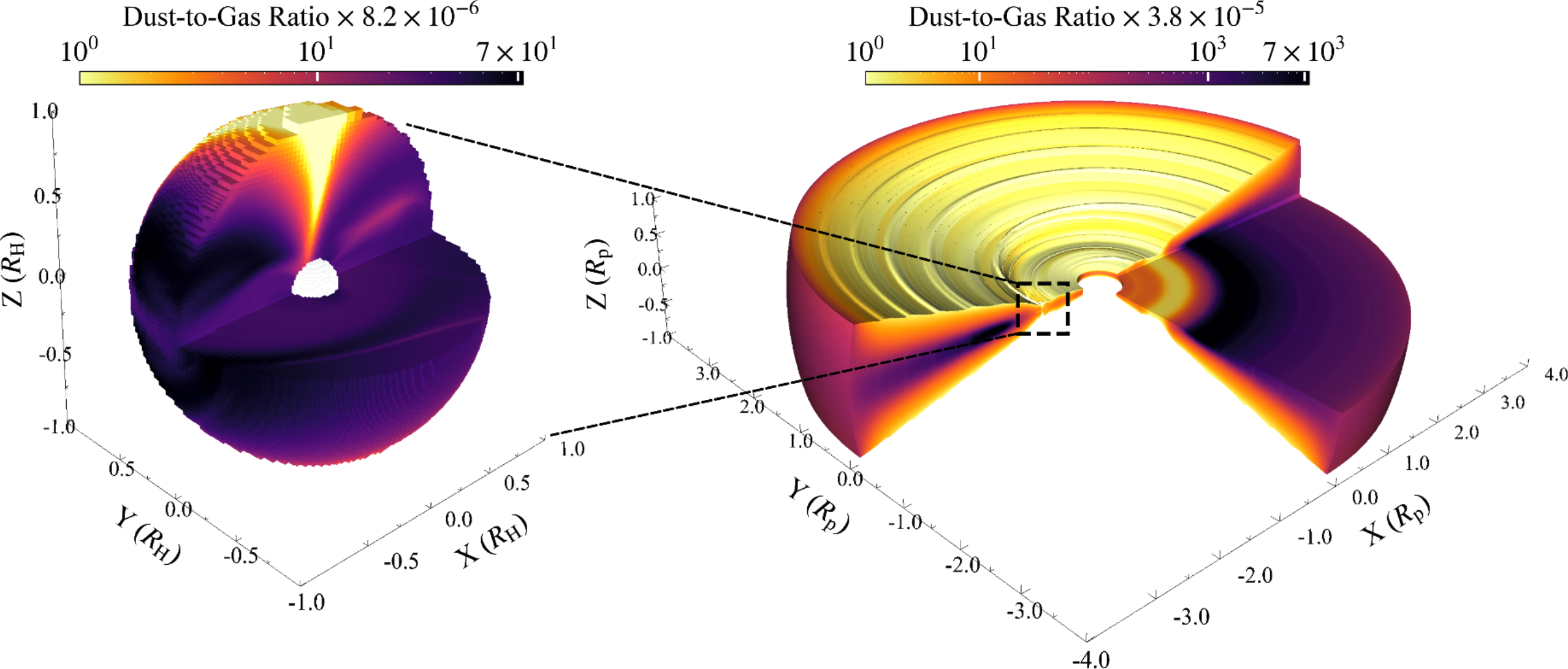}
        \label{fig:sub3_2p5}
    \end{subfigure}
    \caption{The same as Figure \ref{fig:combined}, but now for the case of the 2.5 $M_\mathrm{J}$ planet.}
    \label{fig:combined_2p5}
\end{figure*}

In Figure \ref{fig:combined_2p5}, we show the 3D rendering of the 2.5 $M_\mathrm{J}$ simulation.
The gap is deeper and wider than in the 1 $M_\mathrm{J}$ case causing a sharper transition between the CPD region and the gap within $R_\mathrm{H}$.
The dust-to-gas mass ratio within $R_\mathrm{H}$ is also reduced given the more efficient trapping of the 2.5 $M_\mathrm{J}$ planet.

\begin{figure*}[t]
    \centering
    \includegraphics[width=1.0\textwidth]{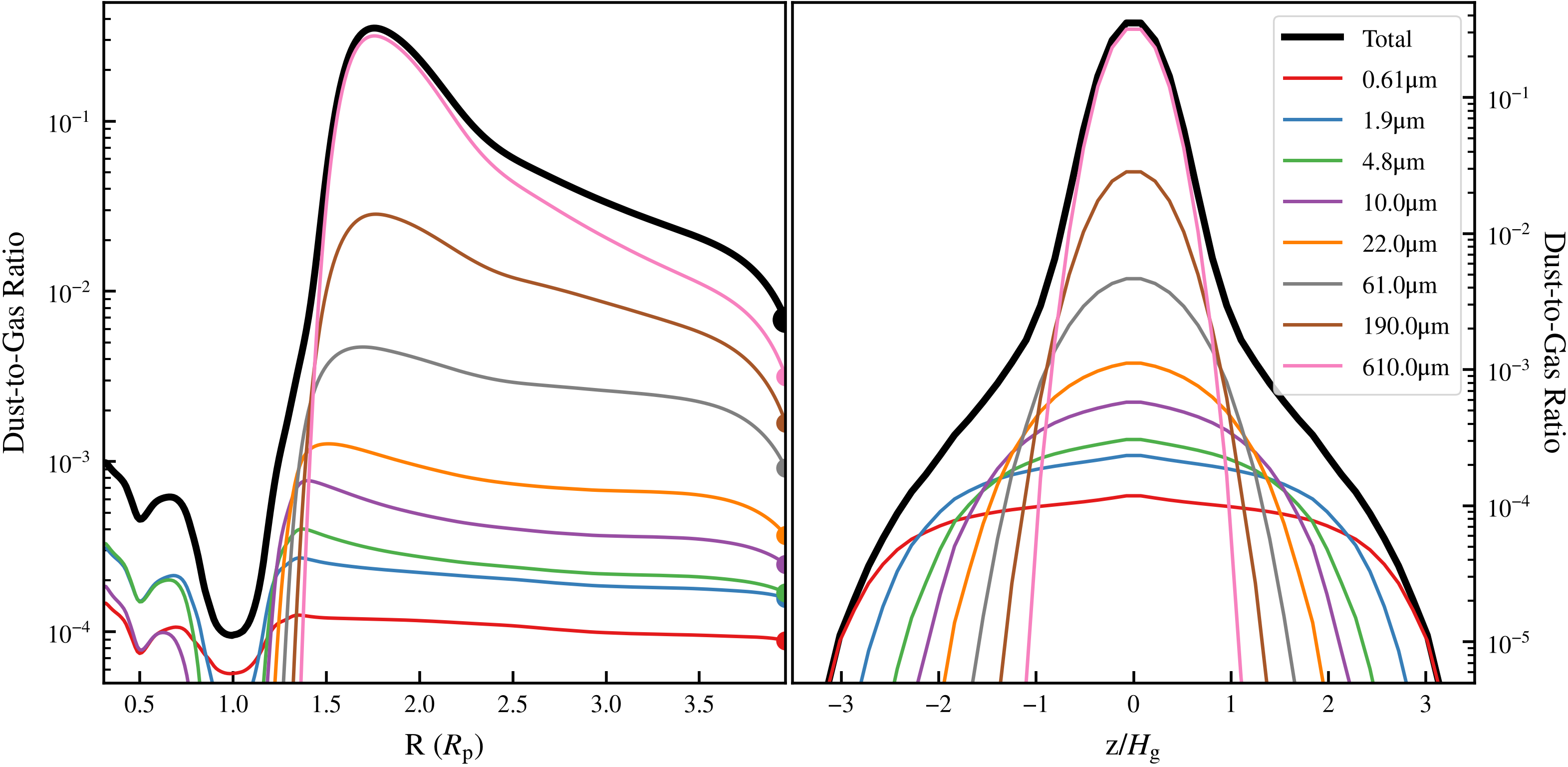}
    \caption{The same as Figure \ref{fig:radial_D2G}, but now for the case of the 2.5 $M_\mathrm{J}$ planet.
    }
    \label{fig:radial_D2G_2p5}
\end{figure*}

In the left panel of Figure \ref{fig:radial_D2G_2p5}, we show the radial dust-to-gas mass ratio in the midplane of the 2.5 $M_\mathrm{J}$ simulation.
In the right panel, we show the dust-to-gas mass ratio as a function of the height above the midplane at the location of the pressure maximum in the disk.
The plot follows the same trends as Figure \ref{fig:radial_D2G}, but there are slight differences between the two simulations.
In the 2.5 $M_\mathrm{J}$ simulation the peak of the midplane dust-to-gas mass ratio occurs at $\sim1.75$ $R_\mathrm{p}$ rather than the $\sim1.5$ $R_\mathrm{p}$ of the 1 $M_\mathrm{J}$ simulation.
Clearly, the more massive planet has moved the peak to a larger radius.
This is also seen in the radial extent of the dust trap which now runs from 1.6 to 2 $R_\mathrm{p}$ rather than 1.4 to 1.9.
The maximum of the dust-to-gas mass ratio is lower in the 2.5 $M_\mathrm{J}$ simulation with a maximum value of $\sim0.35$.
This reduction comes from a higher gas density in the outer disk.

\begin{figure*}
    \includegraphics[width=1.0\textwidth]{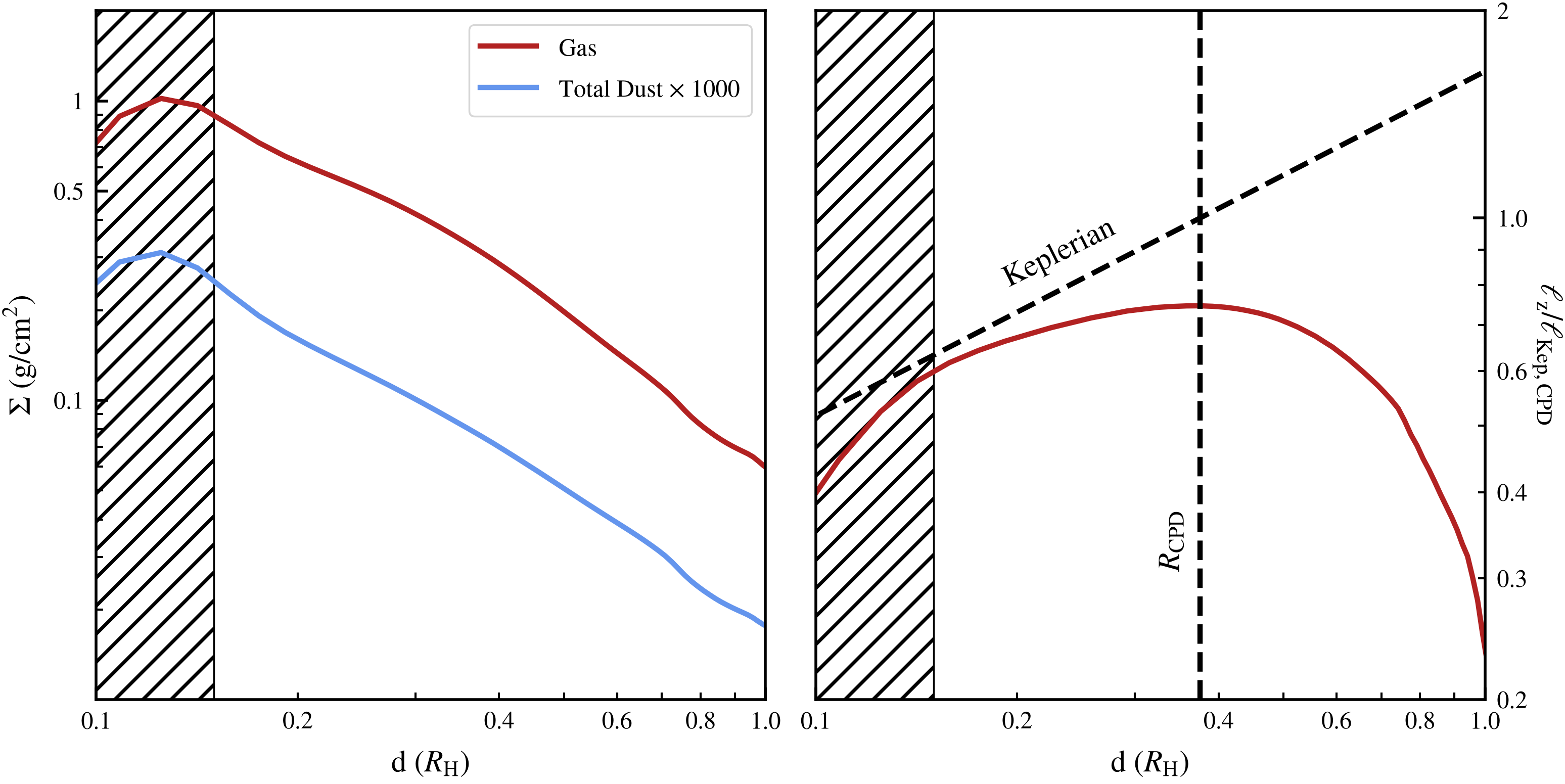}
    \caption{The same as Figure \ref{fig:CPD}, but now for the case of the 2.5 $M_\mathrm{J}$ planet.}
    \label{fig:CPD_2p5}
\end{figure*}

The left panel of Figure \ref{fig:CPD_2p5} shows the azimuthally averaged gas and dust surface densities within $R_\mathrm{H}$, but now for the 2.5 $M_\mathrm{J}$ planet. 
The total dust surface density is scaled by a factor of 1000 to compare with the gas, but even then, the dust is about 1/3 of the gas surface density, indicating a CPD dust-to-gas mass ratio of $\sim0.3\%$.
We also measure the mass from the sink radius to 1/3 $R_\mathrm{H}$, obtaining a gas mass of $1.9\times10^{-4}$ $M_\mathrm{J}$ and a total dust mass of $4.9\times10^{-8}$ $M_\mathrm{J}$, again showing a dust-to-gas mass ratio within the CPD of $\sim0.03\%$.
The right panel of Figure \ref{fig:CPD_2p5} shows the azimuthally averaged specific angular momentum around the planet and is qualitatively the same as the right panel of Figure \ref{fig:CPD}.
Again we define the CPD radius as the point where this specific angular momentum turns over, and find a value of 0.38$R_\mathrm{H}$.

\begin{figure*}
    \centering
    \includegraphics[width=1.0\textwidth]{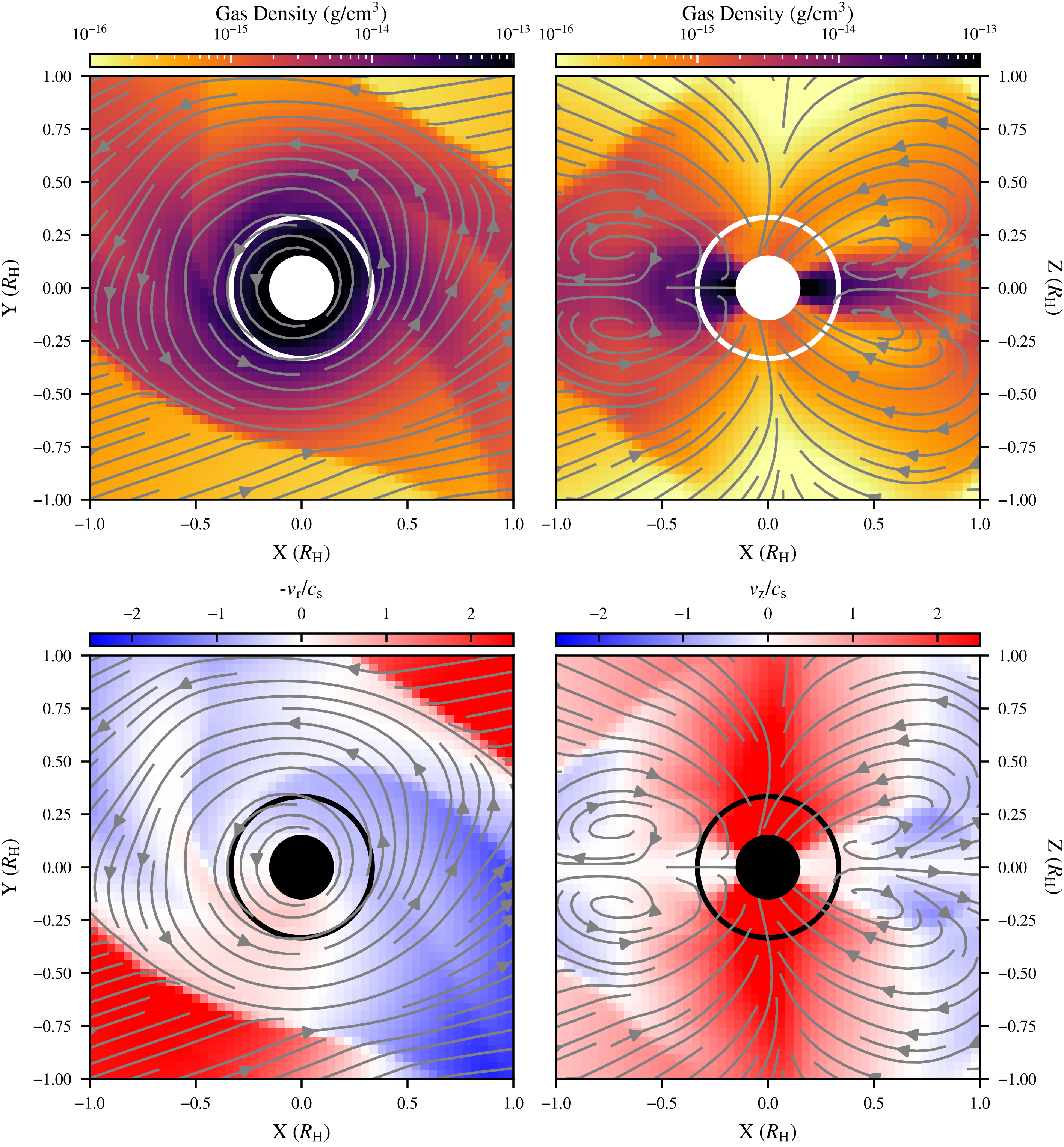}
    \caption{The same as Figure \ref{fig:streamlines}, but now for the case of the 2.5 $M_\mathrm{J}$ planet.}
    \label{fig:streamlines_2p5}
\end{figure*}

The top row of Figure \ref{fig:streamlines_2p5} shows the gas density within 1 $R_\mathrm{H}$ of the 2.5 $M_\mathrm{J}$ planet with associated gas velocity streamlines.
The top left panel shows a midplane cut while the top right panel shows an XZ cut along $\phi=\pi$.
The bottom row shows components of the velocity, with the left being the negative radial velocity so that positive values are directed toward the planet, and the right being the vertical velocity directed toward the midplane.
Compared to Figure \ref{fig:streamlines}, most things are qualitatively the same save for the effects of the increased gravity of the planet.
The midplane density shows more pronounced spiral arms while the vertical extent of the disk is flatter than the 1 $M_\mathrm{J}$ disk.
For the 2.5 $M_\mathrm{J}$ CPD, we measure an aspect ratio of 0.25 at 1/3 $R_\mathrm{H}$ as opposed to 0.36 in the 1 $M_\mathrm{J}$ simulation.
The measured velocities are also more extreme near the planet, both directed toward, and moving away from it.
For example, the maximum vertical velocity achieved is nearly 4 times the sound speed as opposed to 3 times in the 1 $M_\mathrm{J}$ simulation.

\begin{figure*}[!t]
    \centering 
    \includegraphics[width=1.0\textwidth]{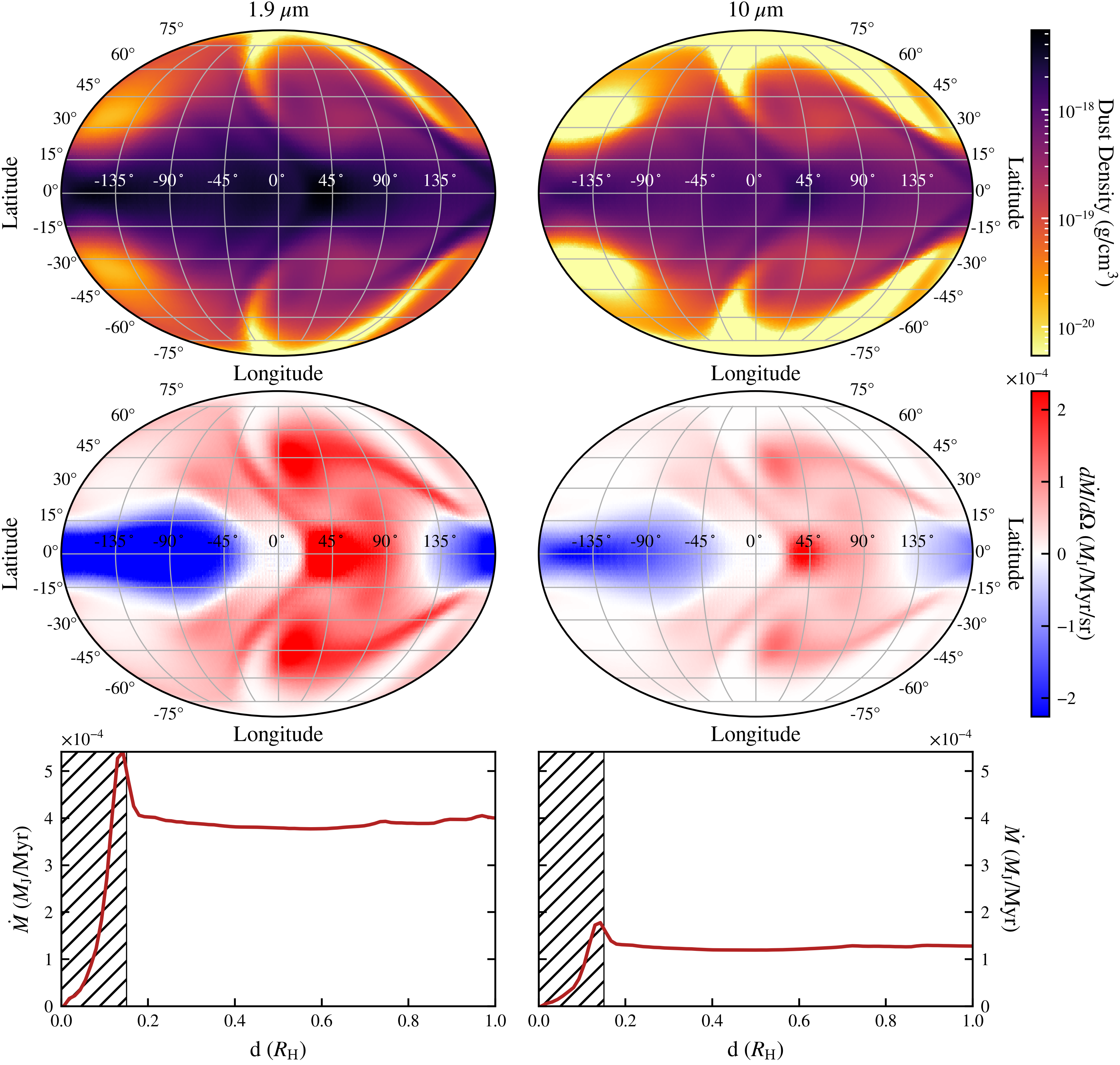}
    \caption{The same as Figure \ref{fig:Money}, but now for the case of the 2.5 $M_\mathrm{J}$ planet.}
    \label{fig:Money_2p5}
\end{figure*}

Figure \ref{fig:Money_2p5} shows the dust density (top row) and mass accretion rate per unit solid angle (middle row) at 1/3 $R_\mathrm{H}$ as well as the mass accretion rate as a function of distance from the planet (bottom row).
The dust densities follow the same pattern with dust size as seen in Figure \ref{fig:Money}, except now the depletion at the poles is more pronounced for smaller sized grains given the stronger filtration effect of the more massive planet (Figure \ref{fig:Filtration}).
The mass accretion rate also exhibits a similar morphology as the middle row of Figure \ref{fig:Money}, with the same prominent midlatitue hot spots and the large removal via the spiral arm.
There is more area in this 1.9 $\mu$m mass accretion panel that has values near 0, again due to the stronger filtration and gravitational influence of the more massive planet.
Like Figure \ref{fig:Money}, the bottom row shows near constant mass accretion throughout $R_\mathrm{H}$ and values at 1/3 $R_\mathrm{H}$ that are equivalent to what is being removed by the sink.

\begin{figure*}
     \centering
    \includegraphics[width=1.0\textwidth]{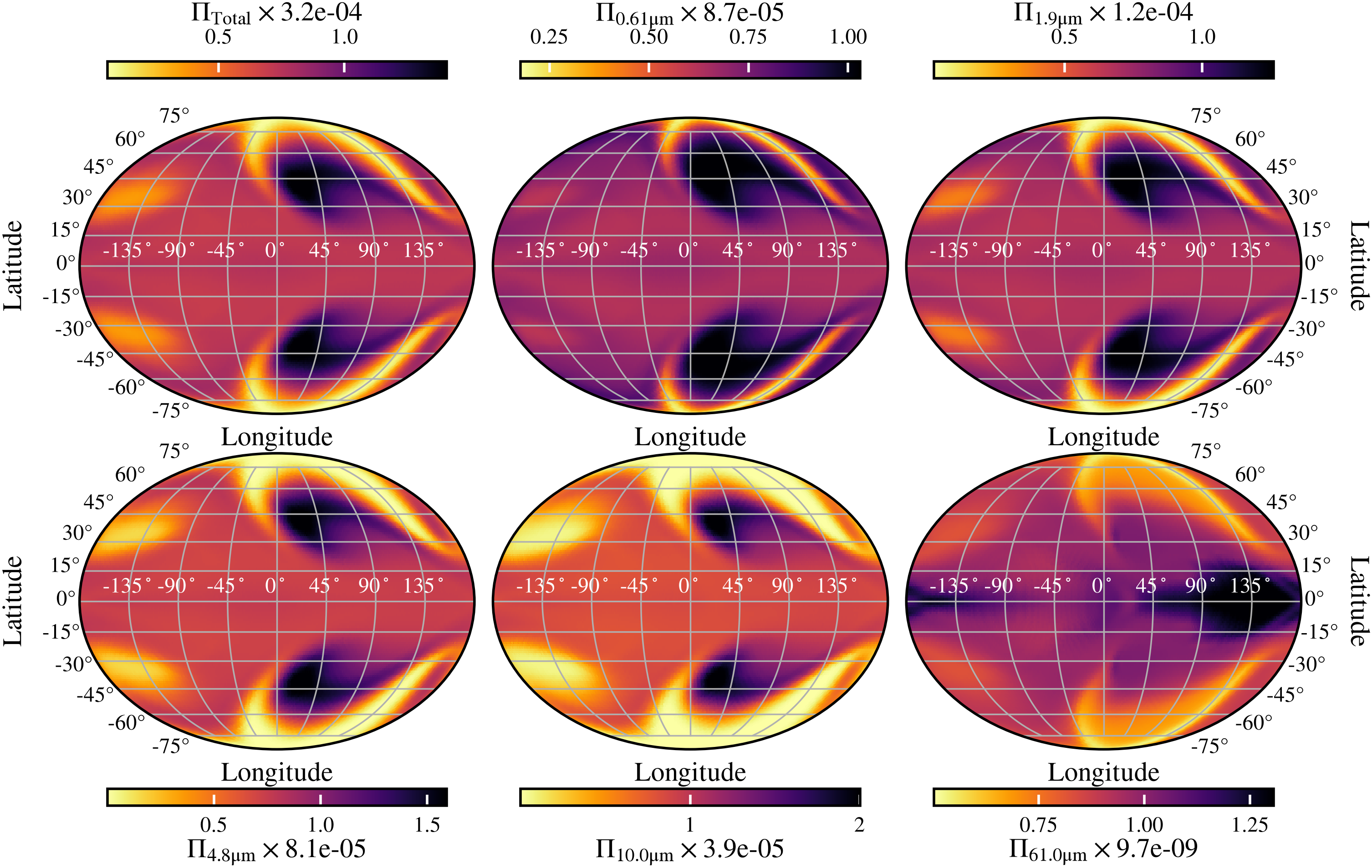}
    \caption{The same as Figure \ref{fig:Mollweide}, but now for the case of the 2.5 $M_\mathrm{J}$ planet.}
    \label{fig:Mollweide_2p5}
\end{figure*}

Figure \ref{fig:Mollweide_2p5} shows Mollweide projections of the dust-to-gas mass ratio at 1/3 $R_\mathrm{H}$, normalized by the average dust-to-gas mass ratio of the respective dust grain at 1/3 $R_\mathrm{H}$.
The figure shows the same pattern seen in Figure \ref{fig:Mollweide}.
In this pattern, smaller grains that supply an appreciable amount of mass to the CPD have hot spots at mid latitudes away from the midplane.
The poles have a reduced dust-to-gas mass ratio.
Eventually, when the grains are large enough, the midplane shows a high dust-to-gas mass ratio as seen in the 61 $\mu$m plot.
Though again, we emphasize that the 61 $\mu$m grain provides a negligible amount of mass to the overall accretion rate.
We exclude the 190 $\mu$m and 610 $\mu$m grains from this plot because they essentially contribute no mass and resemble the 61 $\mu$m.
We also exclude the 22 $\mu$m grain because it resembles the 10 $\mu$m grain.

\clearpage
\newpage
\bibliography{main}{}
\bibliographystyle{aasjournalv7}

\end{document}